# Chemo-mechanical modeling of artificially and naturally bonded soils

Alessandro Gajo[1], Francesco Cecinato[1,*], Tomasz Hueckel[2]

[1]University of Trento, Department of Civil, Environmental and Mechanical Engineering, Trento, Italy.

[2]Duke University, Civil and Environmental Engineering, Durham, NC, USA.

**Abstract**

Chemo-mechanical effects are known to be significant in a number of applications in modern geomechanics, ranging from slope stability assessment to soil improvement and CO2 sequestration. This work focuses on coupled chemo-mechanical modeling of bonded geomaterials undergoing either mechanical strengthening, due to increased cementation, or weakening, due to cement dissolution. A constitutive model is developed that accounts for the multi-scale nature of the chemo-mechanical problem, introducing some cross-scale functions establishing a relationship between the evolution of microscopic variables and the macroscopic material behavior, realistically following the evolution of the reactive surface area, cross-sectional area and the number of bonds along with dissolution/deposition. The model presented here builds up on a previously introduced framework. However, at variance with existing works, it is specialized on materials with only reactive bonds, such as carbonate cemented sandstone or microbially cemented silica sand. Model validation is provided upon reproducing different types of chemo-mechanical experimental datasets, on different naturally and artificially cemented materials, to establish the reliability of the proposed framework.

1. **Introduction**

The presence of some degree of cementation is well known to enhance the strength and stiffness properties of soils, in that it offers an additional contribution in resisting shear and deformation to that provided by an unbonded solid skeleton. For this reason, different soil improvement techniques exist to provide artificial cementation to originally loose or weakly cemented soils; one of the most recently emerged being microbially induced calcite precipitation (e.g. [1]-[4]) to manage mechanical, hydraulic or heat transfer properties [5].

On the other hand, naturally cemented soils or soft rocks may suffer from destructuration, i.e. the more or less gradual degradation and rupture of bonds, for either chemical or mechanical, or combined reasons, bringing about mechanical weakening. Typical examples of such phenomena are landslides triggered by weathering of intergranular bonds, as first discussed by Terzaghi [6] and more recently brought to attention (e.g.) by Zhao et al. [7], in the context of enhanced weathering of soil in landslide slip planes due to acid rain. Several other examples concern slope (often bordering hydroelectric reservoirs), mine pillars instability, etc., due to chemo-mechanical couplings affecting a wide range of materials, from clay [8], [9] to sandstone [10], calcite [11], silica [12], and gypsum [13], just to mention a few.

---

*Corresponding author. Present address: School of Energy, Geoscience, Infrastructure and Society, Heriot-Watt University, Edinburgh, UK. Email: F.Cecinato@hw.ac.uk

Another field where chemo-mechanical effects are seen as potentially important is that of geological sequestration of CO2, an energy generation byproduct. In fact CO2, upon being injected, reacts with brine producing a weak acid, which can in turn react with the surrounding geomaterial, especially if the host reservoir rock is carbonate dominated in composition (e.g. limestone, see [14] and [15]), or is carbonate-cemented (e.g. calcite cemented sandstone, see [16]). Further understanding of chemo-hydro-mechanical couplings in geomaterials is also desirable in the field of unconventional hydrocarbon engineering, as artificial weathering in the form of 'acid stimulation' is often performed to increase permeability of the host rock, typically in carbonate rock or sandstone [17].

In all of the above cited chemo-mechanical phenomena, the hydro-mechanical properties of geomaterials are influenced by dissolution or deposition of bond minerals, mainly calcium carbonate or silicates from and into intergranular bonds. Of particular interest for the safety assessment of different geomechanical applications is bond dissolution, especially in presence of acidic water (e.g., [18] and [19]). In addition to strength, also elastic deformability may be affected by the growth or reduction of cementation bonds, implying a dependency of macroscopic elasticity on plastic deformation, so-called elasto-plastic coupling [20], [21], or plasticity dependent elastic anisotropy (if the process induces elastic anisotropy, see [22], [23], [24] and [25]).

Schematically, based on the kinetics of chemical reaction between different minerals, two main cases can be distinguished, namely the case of both reactive grains and reactive bonds (e.g. calcarenite, with fully carbonate solid skeleton) and the case of non-reactive grains and reactive bonds (e.g. silica sand with carbonate bonds). In the former case, the timescales of dissolution/deposition of grain and bond minerals are comparable, whereas in the latter case the timescale of reaction of the grain minerals is orders of magnitude larger compared to that of bond minerals. As such two cases imply a different overall behavior both at the micro- and at the macro-scale, bespoke modelling is required to address each situation. While in Gajo et al. [26] a model was presented focused on the chemo-mechanical behavior of materials with both chemically reacting grain and bond minerals, this work further develops Gajo et al.'s [26] framework, building up on the cross-scale relationships and making it suitable to capture the chemo-mechanical behavior of bonded geomaterials with only reactive bond minerals and chemically inert grains.

Published experimental data on the loss/gain of strength of geomaterials due to chemo-mechanical effects are not particularly abundant. An account of the phenomenological and modeling background relevant to materials with both reactive grains and bonds, such as calcarenite, was given in [26] (see Section 2 therein), whereas here we review the published work on materials with chemically inert grains, but reactive bonds.

Artificially cemented sand is a primary example of such soils. In the context of soil improvement, DeJong et al. [1] and Montoya and Dejong [27] carried out a series of drained and undrained shear tests on sand cemented by means of microbially induced calcite precipitation. Starting from an initially loose sand, different cementation levels were achieved by subjecting samples to repeated treatments, ranging from lightly cemented to sandstone-like material. The engineered cemented soil was thus shown to undergo during shear a transition from strain hardening to strain softening behavior, in parallel with a significant increase of both stiffness and peak strength.

The behavior of artificially cemented sands was also experimentally investigated by Castellanza [28] and Castellanza and Nova [29], using a different technique. They prepared specimens of silica sand mixed with hydraulic lime and left to cure for a sufficient time to achieve a soft rock-like material. Specimens were then subjected to different chemo-mechanical loading paths in an oedometer apparatus, upon subjecting the material to acid weathering during mechanical testing. Chemical debonding was thus experimentally induced, and it was shown to bring about a significant deformability increase and a decrease of the elastic behavior domain.

Concerning naturally bonded geomaterials, in the context of degradation of mechanical properties due to acidic rain-induced weathering, Ning et al. [30] presented experimental data about calcite cemented 'arkosic' sandstone, subjected to acid treatment (upon immersion of samples into acid solutions of different pH and for different time periods) and subsequently tested in uniaxial compression at different weathering levels. Results demonstrate a clear degradation of both uniaxial compressive strength (UCS) and of elastic modulus with increasing weathering time and decreasing pH of the weathering solution. Also Wang et al. [31] followed a similar weathering procedure on samples of Chinese red sandstone, and subjected them to triaxial compression testing with analogous results, in terms of decrease of peak strength and of elastic modulus with increased weathering.

Several datasets have been published also in the context of $CO_2$ injection-related degradation of natural host rock, with particular regards to sandstone and limestone. For example, Le Guen et al. [32] performed $CO_2$ injection experiments under constant mechanical load in both limestone and sandstone using flow-through triaxial cells, observing a weathering induced increase in strain rates by up to a factor of 5 for limestone, and one order of magnitude smaller for sandstone. This corroborates the above discussed difference in chemo-mechanical behavior between materials with both reacting grains and bonds (limestone) and those with only reacting bonds (sandstone). Further, Hangx et al. [33] interestingly report that for samples of poorly consolidated, carbonate- and quartz-cemented, mainly quarzitic Captain Sandstone subject to brine and $CO_2$ injections during triaxial creep experiments, carbonate bonds dissolved, while quartz bonds remained intact. Despite the complete dissolution of carbonates, no visible change to rock strength was observed. Also Canal et al. [16] performed flow-through triaxial experiments to investigate the effect of $CO_2$ injection on the Corvio sandstone, however in this case the material was reportedly selected in order to minimize chemical couplings. In fact, the Corvio sandstone is a quartzarenite with silica cement, and is consistently shown to undergo minor hydro-mechanical effects due to reactive processes.

In addition, some progress in experimentally studying the micro-scale multi-physics evolution of chemically bonded soils has been recently made [34], [35], [36], [37], [38]. However, quantification of that evolution requires a prior micro-scale model with its well-defined variables, and its quantitative upscaling procedure to allow a macro-scale verification. This is the main purpose of the present work.

From the constitutive point of view, the mechanical behavior of bonded soils/soft rocks has been investigated in the framework of critical state through extended versions of Cam Clay (e.g. see [34], [40]). Based on these works, account for the coupled chemo-mechanical behavior was later taken, in the constitutive models proposed by [41], [28], [29]. However, these works are based on purely macroscopic considerations. Models accounting for the multi-scale nature of the chemo-mechanical problem have been more recently proposed by [17], [11], [27], [42], [43] and [1]. Such micro-to macro scale modeling frameworks have been shown to be generally able to reproduce the experimental behavior of materials with both reactive grains and bonds, especially when they are of the same mineral, with particular reference to calcarenite. However, there are certain peculiarities of behavior of materials with nonreactive grains and reactive bonds and some advantages of their targeted modeling as such.

In this work, a microscale inspired chemo-mechanical modelling framework specialized on materials with only reactive bonds is presented, building up on the model introduced in [1], and validated against different types of experimental chemo-mechanical loading paths. Differently from most of the existing chemo-mechanical models, reversible behavior is described here in the framework of hyperelasticity, upon defining a suitable elastic free energy, allowing for the occurrence of destructuration through the dependency of elastic properties on chemo-mechanical plastic straining (elasto-plastic coupling, cf. [20], [21]). Notably, two key macroscopic 'cross-scale' functions are introduced, establishing a relationship between the evolution of microscopic variables and the macroscopic material behavior, with the benefit of avoiding the occurrence of unphysical situations, such as uninhibited deposition at no available pore volume, or continuing dissolution when no cement mineral is left. The model allows to follow the evolution of the reactive surface area, cross-sectional area and the number of bonds along with dissolution/deposition in a realistic, yet simplified, manner, and to correctly reproduce the macroscopic chemo-mechanical behavior of cemented materials with only reactive bonds subjected to experimental testing. In summary, compared to previous works, with particular reference to [1], the main innovative aspects of this work include (i) the capability to deal with materials with nonreactive grains, (ii) validation against experimental tests involving not only chemo-mechanical dissolution/destructuration, but also cement augmentation leading to mechanical strengthening and stiffening of the material, (iii) a more in-depth examination of certain key aspects of the framework introduced in [1], such as the effect of cement deposition on elasticity and the asymptotic behavior of the specific reactive surface area close to null porosity, and (iv) a simpler and more didactic approach where the key aspects of the proposed dual scale model are clearly illustrated, and even summarized in a dedicated section (2.3.1), to facilitate understanding of the model structure.

The paper is structured as follows: in Section 2 the main aspects of the model's macroscopic and microscopic formulation and calibration are described, in Section 3 the model validation against experimental data on both artificially and naturally bonded geomaterials is discussed, and conclusions are drawn in Section 4.

## 2. Model description

In this section, the main features of the model formulation and calibration are described. The key relationships linking the micro- and the macro-scale consist of a set of phenomenological cross-scale functions (Section 2.3), which are subsequently calibrated on a REV composed of eight non-reactive grains (in contact or not, with one to another), where each grain can be connected to up to four other grains by means of mono-mineral bonds (cf. Section 2.4). However, the cross-scale functions are not tied to a specific microscopic geometry; in fact they could be adapted to different REV configurations without affecting the model's core assumptions and predictive capabilities.

### 2.1 Macro-scale model description

#### 2.1.1 Chemo-elastic behavior

Additive decomposition into elastic and plastic parts is assumed for the macroscopic strain tensor. Each elastic and plastic strain include a chemical component, proportional to a mineral mass loss/gain, in this case of intergranular bonds. The elastic behavior of the material can be deduced by defining a suitable elastic free energy density function, in the framework of hyperelasticity and elasto-plastic coupling [20], [21]. The standard form of free energy for the unbonded soil $\varphi_g = \varphi_g(\boldsymbol{\varepsilon}_e)$ is modified, to account for the presence of both mechanically and chemically interacting bonds, into

$$\varphi = \varphi(\boldsymbol{\varepsilon}_e, \boldsymbol{\varepsilon}_p, m_b) \qquad (1)$$

where $m_b$ is the (time-integrated) mass change of all mechanically active cementing bonds per unit volume, from a reference configuration $m_{b0}$ (see Section 2.2.1). For simplicity, a detailed description is provided here only for chemo-mechanical elastic weakening of the material, i.e. only the role of mechanical destructuration and chemical dissolution effects in the material elastic behavior are discussed. The more complex effects of chemical deposition, causing elastic strengthening, are discussed in Appendix A.

The macroscopic elastic free energy is postulated as linearly interpolated between the energy of cementation-free configuration and that of totally cemented configuration

$$\varphi(\boldsymbol{\varepsilon}_e, \boldsymbol{\varepsilon}_p, m_b) = (1-\tilde{\omega})\, \varphi_g(\boldsymbol{\varepsilon}_e) + \tilde{\omega}\, \varphi_b(\boldsymbol{\varepsilon}_e) \qquad (2)$$

where $\varphi_g$ is the free energy of the uncemented solid skeleton, and $\varphi_b$ is the free energy of the completely cemented soil (implying a larger stiffness than that of the unbonded material, since the whole porous space is filled with cement).

The weighting function $0 \leq \tilde{\omega} \leq 1$ is set to be a function of a microscopic variable, namely the mean specific cross section area of mechanically active cementing bonds $a_b$ (that depends on both cement mass change and plastic strain, see Section 2.3), as $\tilde{\omega} = a_b^\alpha$, where $\alpha$ is a positive constitutive parameter. Note that both free energies depend on the same variable of macro-scale elastic strain.

It is worth adding that eqn. ( 2) is not formulated within the rigorous upper and lower bounds framework, as proposed by Voight [43] and Reuss [45] because these bounds would not take account of the rupture of cementation bonds. Voight and Reuss bounds could, however, be used for the evaluation of the free energy of completely cemented soil $\varphi_b$, which can be expressed as the weighted volumetric average of the free energies of the minerals constituting grains and bonds.

Mechanically induced destructuration is quantified by plastic strain and describes a reduction of the number of mechanically active bonds, represented by the decrease of their mean specific cross section area $a_b$, and consequently of the weighting function $\tilde{\omega} = a_b^\alpha$. This implies that the first right hand term of eqn. ( 2) will gradually acquire larger importance compared to the second term, as destructuration progresses, causing ultimately an overall material stiffness decrease. At the limit of completely unbonded material (i.e. when all the active cement bonds are broken, or dissolved, $a_b = 0$) the free energy of the uncemented solid skeleton $\varphi_g$ is recovered. On the other hand, at the limit $a_b \to 1$, the free energy becomes that of completely cemented material $\varphi_b$. In other terms, chemically induced cement dissolution causes a fraction of bonds (generally carrying a fraction of the overall stress) to disappear (i.e., reducing $a_b$), leading in turn to a stress increase in the remaining bond fraction, inducing strains at constant stress. Thus, cement dissolution causes a decrease of soil stiffness with associated strains (if dissolution occurs at constant stress) or a decrease of soil stiffness with an associated decrease of applied stress (if dissolution occurs at constant strain). Eqn. ( 2) consistently accounts for these effects. On the other hand, cement deposition induces an increase of material stiffness, at constant stress and strain. The above described behavior is taken into account with a step-wise form of the elastic energy, presented in Appendix A.

It should be observed that in eqn. ( 2) only stiffness degradation due to bond breakage is considered, while the increase of stiffness due to compaction in a destructured material is neglect for simplicity.

### 2.1.2 Plastic behavior

In the framework of small strains, assuming the solid grains incompressible and an isotropic material behavior, we propose a yield function along the lines of Modified Cam Clay, extended to account for an isotropic tensile strength function (e.g. see [34], [46], [41] and [1]), as follows:

$$F\left(\boldsymbol{\sigma}, \boldsymbol{\varepsilon}_p, a_b\right) = \left[M_{cv}L(\theta)\right]^2 \left[(p+p_{tens})^2 - (p+p_{tens})(p_c + p_{comp} + p_{tens})\right] + q^2 \quad (3)$$

where $M_{cv}$ is the critical state parameter, $L(\theta)$ is a function of the Lode angle describing the deviatoric section of the yield surface (see [47]), $p$ the mean effective pressure, $q$ the deviatoric stress, $p_c$ is the preconsolidation pressure of the uncemented soil, and $p_{tens}$ and $p_{comp}$ respectively represent the increase of tensile and compressive strength, compared to uncemented soil, due to cementation. It is assumed that both quantities depend on the mean specific cross section area of all mechanically active bonds $a_b$, as

$$p_{tens} = a_b \sigma_{rt} \quad \text{and} \quad p_{comp} = a_b \sigma_{rc}. \quad (4)$$

In the above, $\sigma_{rt}$ and $\sigma_{rc}$ are the macroscopic uniaxial tensile and compressive strength of the material constituting the cementing bonds. Since typically $\sigma_{rt}$ can be taken as a fraction of $\sigma_{rc}$ (see Sect. 3), $p_{tens}$ and $p_{comp}$ turn out to be proportional to one another. It can be observed that when all cementation bonds are broken or inexistent, $a_b = 0$, thus equation (3) becomes that of the Modified Cam Clay yield locus. Figure 1 shows the evolution of the yield surface with cementation/destructuration, in terms of invariants $(p, q)$.

The preconsolidation pressure $p_c$ is expressed as

$$p_c = p_0 \exp\left\{\frac{1+e_0}{\lambda}\left[\varepsilon_p^v - \varepsilon_0\right]\right\} \quad (5)$$

where $p_0$, $\varepsilon_0$ and $e_0$ the reference mean pressure, volumetric strain and void ratio respectively, $\varepsilon_p^v$ the plastic volumetric strain and $\lambda$ a constitutive parameter. Equation (5) represents a hardening relationship which has been simplified, compared to standard Cam Clay. A more complete discussion on the hardening relationship in the case of deposition/dissolution is provided in Appendix B. It is worth emphasizing that the hardening relationship (5) has been selected for the sake of simplicity in the presentation of the model, however, the proposed constitutive framework is not limited to such assumptions. Equation (5) represents the hardening state of the material in its unbonded condition, i.e. when $p_{tens} = 0$ and $p_{comp} = 0$, and hence depending on the accumulated volumetric plastic strain $\varepsilon_p^v$ only. However, in a generic situation, couplings are introduced in equation (3) through functions $p_{tens}$ and $p_{comp}$, which depend on the amount of mechanically active bonds (eqn. (4)), thus ultimately depending on both plastic strain and on mass deposition/dissolution, as is shown in Section 2.3.

It is worth remarking that the above described plastic driver was adopted due to its simplicity. However, to better reproduce the behavior of certain sandy materials, modifications to relevant constitutive ingredients (e.g., yield function, hardening rule and elastic response) could be carried out without substantially altering the proposed framework. In particular, a more suitable hardening rule would not only involve isotropic hardening of the yield surface (resulting from cementation/decementation), but also a change of its shape, so that the behavior of the uncemented material would be represented by a conical yield surface with a cap.

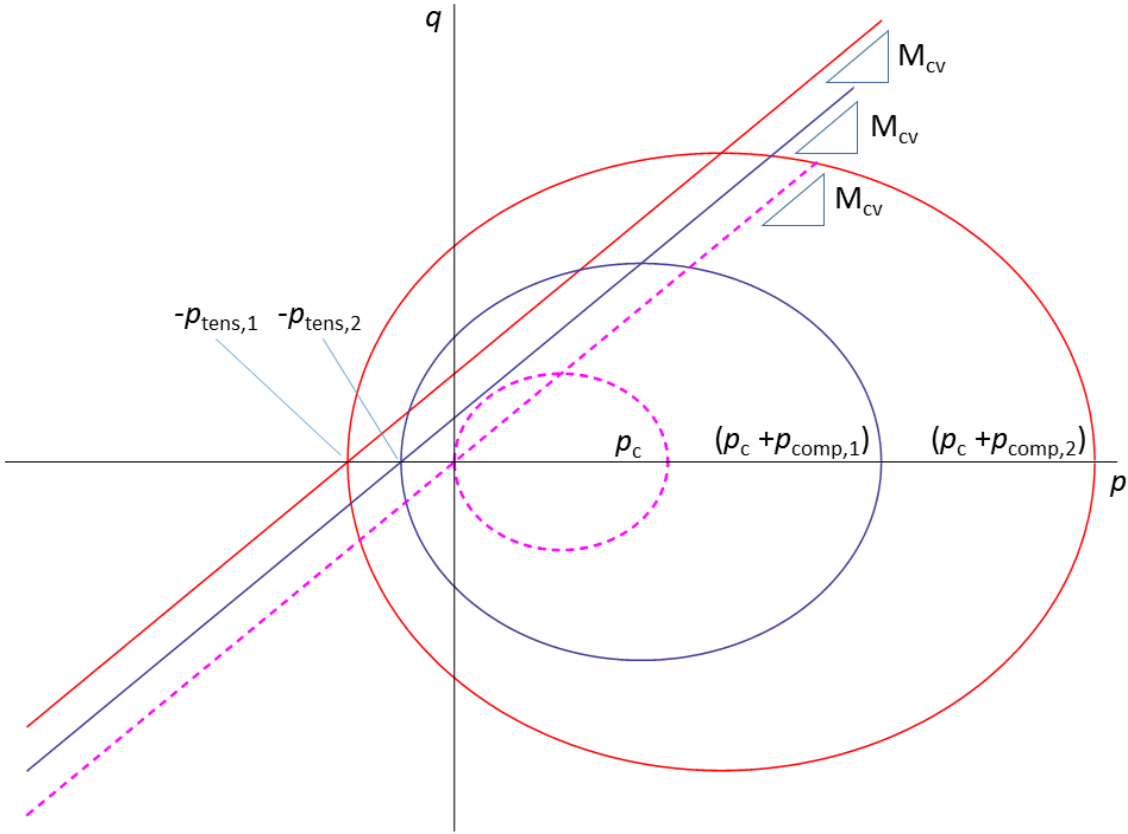

Figure 1. Change of size of yield surface with changing cementation/destructuration in terms of invariants (p, q). The dashed surface represents the uncemented state, in which $p_{tens}=p_{comp}=0$.

## 2.2 Rate equations
### 2.2.1 Chemical Kinetics

Macroscopically, chemical kinetics controls the mass change $\dot{m}_b$ within the REV due to dissolution and/or deposition, at a rate quantified by the mineral kinetic equation which is proportional to the specific reactive surface area $a_r$ per unit volume of the reacting materials (with units m²/m³). In the present context, this quantity corresponds to the surface area of the bonds that are actually in contact with the pore fluid (Section 2.3). The chemical reaction rate can be thus described by the following equation expressing the specific mass change (per unit time) of a specific bond mineral 'i' (e.g. [47]):

$$\dot{m}_b^i = a_r^i k^i \left( \mu_s^i - \mu_f \right) \tag{6}$$

where $k^i$ is a rate constant (with units (kg mol)/(m² s J)), $a_r^i$ the specific surface area of the mineral (with units m²/m³) per the same volume, and $\mu_s^i$ and $\mu_f$ are the chemical potentials (with units of J/mol) of the reacting solid mineral and of the fluid respectively. In general, at constant chemical potential difference between the reacting mineral and the fluid, the mass rate of change evolves with the microscopic geometry of the system (represented by $a_r$). The relationships between the specific reactive surface area and the microscopic features of the solid skeleton on the one hand, and the porosity on the other hand, are key features of our model providing coupling between microscopic and the macroscopic description of porous media, and are discussed in detail in Section 2.3. It should be remarked that, for the sake of simplicity, in this work we consider mono-mineral bonds, and open systems (i.e., the dissolved mass is transported away by the fluid), although the presence of more than one family of bonds could be easily introduced. Modelling the transport of species within and outside the REV is beyond the scope of this work.

### 2.2.2 Destructuration equation

In general, considering a REV of bonded geomaterial, it can be observed that not all bonds are mechanically active, due for instance to the rupture of part of the bonds consequent to mechanical loading. Thus, we define $N_b$ (with dimensions of 1/L³) as the number of potential (i.e. both broken and unbroken) bonds per unit volume of the porous skeleton, and $N_{ba}$ as the number of mechanically active (i.e. unbroken) bonds per unit volume. $N_{ba}$ ranges between 0 and $N_b$, with $N_{ba} = 0$ when all bonds are broken, and $N_{ba} = N_b$ when all bonds are mechanically active. Quantity $N_{ba}$ is bound to depend both on mechanical loading (bringing about destructuration) and on the deposition of cement at broken bonds (eventually restoring the integrity of previously broken bonds).

The above described mechanisms of destructuration and cement deposition are reproduced in the following empirical function:

$$\dot{N}_{ba} = -k_1 N_{ba} \sqrt{(1-A)(\dot{\varepsilon}_p^v)^2 + A(\dot{\varepsilon}_p^s)^2} + k_2 (N_b - N_{ba})\langle \dot{m}_b \rangle \quad (7)$$

where $A$, $k_1$ (dimensionless) and $k_2$ (with units m³/kg) are constitutive parameters, with $A$ ranging between 0 and 1 to weigh the role of volumetric ($\dot{\varepsilon}_p^v$) and deviatoric ($\dot{\varepsilon}_p^s$) plastic strain rates in destructuration. The Macaulay brackets in the second addend of eqn. (7) are used since cement deposition only is assumed to form new active bonds, while plastic strains are assumed to induce only bond rupture. Positive $\dot{m}_b$ is associated with bond mass increase.

It can be observed that cement dissolution/deposition $\dot{m}_b$ is expected to affect both porosity and the thickness of bonds. Thus, starting from the standard definition of porosity, in general the rate of porosity change of the material due to chemo-mechanical loading depends on both mass change and on volumetric strain, as follows:

$$\dot{n} = -\frac{\dot{m}_b}{\rho_s} - (1-n)Tr(\dot{\varepsilon}) \quad (8)$$

where $\rho_s$ is constant, since solid grains are considered incompressible. In the framework of small strains theory, it is found appropriate to also introduce a 'chemically affected' porosity $\tilde{n}$, defined as the ratio of current voids volume over the initial bulk volume and depending only on mass dissolution/deposition, through the following rate equation:

$$\dot{\tilde{n}} = -\frac{\dot{m}_b}{\rho_s} \quad . \quad (9)$$

### 2.3 Cross-scale relationships

The constitutive concepts developed so far are based on two key macroscopic quantities, the cross sectional area of active bonds $a_b$ and the reactive surface area $a_r$, which can be defined in terms of microscopic variables. To relate the evolution of microscopic variables with the macroscopic chemo-mechanical description of the material, reference is made to a simplified microscopic geometry.

In Figure 2a, a schematic of the micro-scale geometry, inspired by microscope photographs of thin sections of cemented materials (e.g. calcarenite, cf. [49] and [1]) is shown, in the general case where grains might not be in direct contact, but they are linked by cementation bonds that are assumed to be isotropically distributed. In such large porosity configuration, the geometry of bonds and grains can be approximated respectively with cylinders and spheres. In Figure 2b, the case is represented where, regardless of the size of bonds, grains remain in direct contact, resulting in a simplification of the solids' geometry.

It can be argued that the actual microscopic structure of a cemented material with chemically reacting bonds schematically lies in between the two cases shown in Figure 2. In fact, even if cementation takes place in a soil after its geological deposition, cement may be partly deposited around pre-existing grain contacts, and partly forming bridges between grain areas that were not originally in contact (this is corroborated by observing microscopic photographs of cemented sand, e.g. [1]). Thus, in the mathematical description of geometry at particle level that follows, reference will be made to geometrical parameters representing average quantities, with particular reference to an average intergranular distance $d$.

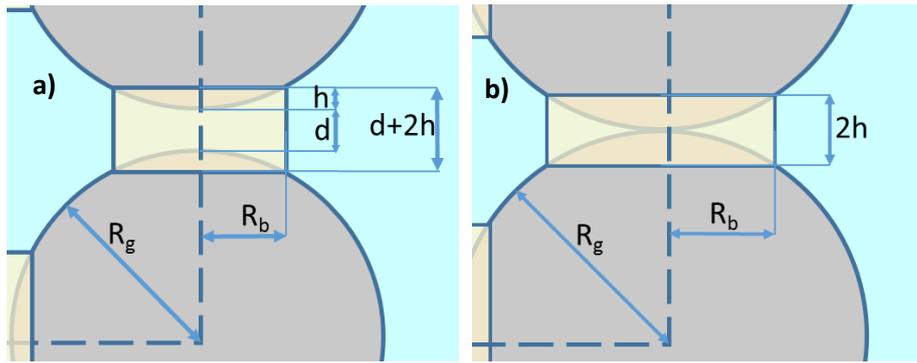

*Figure 2. Schematic of the considered simplified geometry at large porosity, (a) in a general case where the spherical grains are not in direct contact, but they are linked by cylindrical bonds, and (b) in the case where the spherical grains are in direct contact, and bonded by cylindrical bonds.*

Considering a large porosity configuration, with reference to the general case of Figure 2a, we can express the bulk volume of the grains as

$$v_g = \mathsf{N}_g \frac{4}{3}\pi R_g^3 \,, \tag{10}$$

where $\mathsf{N}_g$ the number of grains (per unit initial volume of the porous skeleton) and $R_g$ the mean radius of the grains, which is a constant as the grains are assumed nonreactive. The height of the grain portion wrapped by the bond can be expressed as $h = R_g - \sqrt{R_g^2 - R_b^2}$, where $R_b$ the (variable) mean radius of the bonds, and the mean length of the bonds can be defined as $L_b = d + 2h$, where $d$ the average intergranular distance (i.e. the distance between the edges of two grains connected by a bond).

Thus, the bulk volume of cementing bonds can be deduced as

$$v_b = N_b \pi \left[ R_b L_b^2 - h\left(R_b^2 + h^2/3\right) \right], \quad (11)$$

hence the volume of the solid phase (per unit initial volume of the porous skeleton) coincides with $1-\tilde{n} = v_b + v_g$, where $\tilde{n}$ the 'chemically affected' porosity defined by eqn. (9).

The specific reactive surface area $a_r$ can be expressed as the sum of the lateral surface area of all the bonds and of the cross section areas of the broken bonds (which are $N_b - N_{ba}$), as

$$a_r = N_b 2\pi R_b L_b + (N_b - N_{ba}) \pi R_b^2. \quad (12)$$

Equation (12) represents the limit of large porosity. On the other hand, at low porosity configurations (e.g. when the volume of bonds is large), the specific reactive surface area becomes very small and the schematics of Figure 2 are no longer representative of the microscopic geometry. Thus, in this case the microstructure can be related to the macroscopic behavior in a phenomenological fashion, introducing restrictions to avoid physical inconsistencies. In particular, when the growth of bonds fills all the available porous space, the porosity is prescribed to tend to zero; and when all bonds are dissolved, the specific reactive surface area $a_r$ is imposed to tend to zero. The following general asymptotic relationship between $a_r$ and $n$ in the limit of very small porosity is proposed:

$$a_r = a_{r0} n^\gamma \quad (13)$$

where $a_{r0}$ and the exponent $\gamma$ are constitutive parameters. A demonstration of such constraint is presented in Appendix C.

Combining eqns. (12) and (13) with an appropriate weighting function $\omega_1$, a relationship is obtained expressing the specific reactive surface area of materials with chemically reacting bonds throughout the porosity range:

$$a_r = \left[ N_b 2\pi R_b L_b + (N_b - N_{ba}) \pi R_b^2 \right]\left(1-\omega_1^\beta\right) + a_{r0} n^\gamma \omega_1^\beta \quad (14)$$

where $\beta$ a constitutive parameter and the weighting function $\omega_1$ is expressed as

$$\omega_1 = v_b + v_g. \quad (15)$$

It should be noted that eqn. (15) effectively inhibits the exchange of cement mass (due to dissolution/precipitation) when there are no cementing bonds, or when there is no void volume left. The condition of absent cementing bonds can be reached only asymptotically, as a vanishingly small quantity of cementation (and a vanishingly null pore volume on the other hand) will always exist.

Along the lines of Gajo et al. [1], knowing the number of active bonds per unit volume $N_{ba}$, the number of active bonds per unit initial cross section of the porous skeleton can be approximately estimated as $\tilde{N}_{ba} = N_{ba}^{2/3}$. Thus, the mean specific cross section area of cementing bonds may be expressed at the limit of very large porosity as $a_b = \pi R_b^2 \tilde{N}_{ba}$, and at the limit of small porosity, as $a_b = 1-\tilde{n}$.

Applying similar interpolation concepts to those adopted for the reactive specific area, we obtain the following expression for the specific cross section area of cementing bonds throughout the porosity range:

$$a_b = \pi \aleph_{ba} R_b^2 \left(1-\omega_2\right)^\theta \left(1-\omega_1\right)^\delta + (1-\tilde{n})\omega_2^\theta \omega_1^\delta \qquad (16)$$

where $\delta$ and $\theta$ are constitutive parameters, and $\omega_2$ is a weighting function:

$$\omega_2 = \left(\frac{v_b}{1-v_g}\right). \qquad (17)$$

In the above, (1- $v_g$) represents the bulk volume that is available for a complete cementation of bonds. In fact Eqn. (17)a takes null value when $v_b = 0$ (no cementing bonds), and equals one when the growing bonds fill all the available void space, i.e. for $v_b = 1 - v_g$ (no void volume). Thus, the specific cross section area of cementing bonds $a_b$ ranges between 0 and 1.

The interpolation functions proposed above were deduced from theoretical arguments and from the calibration on simple arrays of spheres and cylinders described in Section 2.4. However, the proposed model is not restricted to the particular form of interpolation functions given in eqns. (14) and (16), thus such functions should be adapted to different micromechanical models.

It should be observed that, under our model's assumptions, a rearrangement of the grains (induced by elastic and/or plastic strains) is not expected to directly affect the cross-scale quantities (i.e. the specific surface area and the specific cross section area of the bonds), unless indirectly, through destructuration. Along these lines, the specific cross section area of the bond has been assumed to depend on $\tilde{n}$ instead of $n$, thus it depends only on chemical dissolution/deposition, without being affected by grain rearrangement (otherwise, an elastic volume compression would imply an increase of cementation). This is consistent with the assumption of small strains, because the effects of cement deposition/dissolution on shear strength are taken into account making reference to the initial, undeformed configuration. This argument certainly applies also to the specific surface area of the bonds. However, since the specific reactive surface area of the bonds has the dual role of ruling the rate of chemical reactions of dissolution/deposition on the one hand, and of preventing unphysical situations (in which further cement may precipitate with no available pore space, or dissolve with no available cement) on the other hand, it was deemed more appropriate to assume that the reactive surface area depends on $n$ (instead of $\tilde{n}$), so that its role of preventing unphysical situations is preserved, although a small error in the evaluation of the rate of chemical reactions is introduced.

### 2.3.1 Summary of model variables and cross-scale functions

To facilitate understanding of the model structure, in Table 1 a list of the main variables introduced above is reported, in which the different quantities have been categorized into micro- and macro-scale quantities, and the main cross-scale functions are listed with their functional dependency.

Following the above outlined framework, the model can be easily extended to more complex cases, e.g. to materials with more than one kind of cementing bonds, with different strengths and kinetics of chemical reaction. For the sake of brevity, such generalized formulation is not reported here.

Table 1. Main introduced variables categorized into micro- and macro-scale quantities, and key cross-scale functions.

| Macro-scale quantities | | Micro-scale quantities | |
|---|---|---|---|
| symbol | name | symbol | name |
| $\boldsymbol{\sigma}$ | effective stress tensor | $L_b$ | mean length of bonds |
| $\boldsymbol{\varepsilon}$ | strain tensor | $d$ | mean distance between grain edges |
| $\sigma_{rt}, \sigma_{rc}$ | tensile and compressive strength of cement | $v_b$ | bond bulk volume |
| $p_c$ | preconsolidation pressure | $R_b$ | mean bond radius |
| $\varepsilon_p^v$ | plastic volumetric strain | **Cross-scale functions** | |
| $\varepsilon_p^s$ | plastic deviatoric strain | equation reference number | function dependencies |
| $a_r$ | specific reactive surface area | (14) | $a_r = f(R_b, L_b, d, v_b, \boldsymbol{\varepsilon})$ |
| $a_b$ | mean specific cross section area | (16) | $a_b = f(R_b, \boldsymbol{\varepsilon}_p, a_r)$ |
| $n$ | porosity | (8) | $\dot{n} = f(\dot{m}_b, \dot{\boldsymbol{\varepsilon}})$ |
| $\dot{m}_b$ | mass change due to deposition/dissolution | (6) | $\dot{m}_b = f(a_r, \mu_s, \mu_f)$ |
| $\mathbb{N}_{ba}$ | change of number of active bonds per unit volume | (7) | $\mathbb{N}_{ba} = f(\dot{m}_b, \dot{\boldsymbol{\varepsilon}}_p)$ |
| $p_{tens}, p_{comp}$ | increase of tensile and compressive strength due to cementation | (4) | $p_{tens}, p_{comp} = f(a_b)$ |

### 2.4 Cross-scale calibration

In this section a calibration of the above outlined cross-scale functions is provided with reference to a sample microscopic array of grains and bonds. A cubical REV is considered, composed of eight spherical grains connected by cylindrical bonds. The spherical grains are centered in the REV's vertices and the cylindrical bond axes lie in the REV's edges, so that the porous space is located at the center of the REV. Microscopic geometrical calculations are carried out leading to expressing the specific bond cross-sectional area $a_b$ and the specific reactive surface area $a_r$ in a rigorous manner throughout the porosity range, without resorting to the phenomenological simplifications described in Section 2.3. The resulting trends obtained with such microscopic representation are then used to calibrate the constitutive parameters featuring in eqns. (14) and (16) through a trial and error procedure, and then compared to those obtained with the cross-scale functions of Section 2.3.

Two reference configurations are considered for the microscopic geometry, namely a REV of identical spherical particles that are not in direct contact (Figure 3a), so that the inter-granular distance $d \neq 0$, and a

REV of identical spherical particles that are in direct contact (Figure 3b) implying $d=0$. The latter configuration is described hereafter, while details about the former can be found in [1]. It should be noted that no bond rupture by mechanical action is considered in these calculations, thus the considered geometry of the solid skeleton changes due to uniform deposition/dissolution of bonds only.

The microscopic representation is subdivided into two geometrical configuration sub-cases: a first configuration at larger porosity, in which bonds undergo uniform thickness reduction/increase. This configuration is lower-bounded by the complete dissolution of bonds when $n \rightarrow 1 - V_g/V$, where $V_g$ the volume of grains and $V$ the total volume of the REV, and upper-bounded at the point when the laterally growing bonds (cylinders) come into contact, corresponding to the geometrical condition $R_b \leq R_g/\sqrt{2}$. Further deposition beyond this point implies a second configuration, at lower porosity, where the skeleton geometry is represented by interpenetrating cylinders. The second configuration is limited by the virtually complete filling of porous space, hence it applies to the range $R_g/\sqrt{2} \leq R_b \leq R_g$.

In Figure 4a the specific cross section area of cementing bonds $a_b$ is plotted versus porosity, considering a microscopic REV with $R_g = 0.05$ mm, at different values of the initial bond radius $R_b$ and of the inter-granular distance $d$ (including $d=0$). The discontinuity in the curves represents the passage between the first and second geometrical configuration. In Figure 4b the corresponding curves obtained with the macroscopic calculation of eqn. (16) are shown. Adequate correspondence between the two sets of curves has been obtained upon parameter calibration, resulting in $\delta = 1.5$ and $\theta = 0.5$.

In Figure 5a, the specific reactive surface area $a_r$ obtained by microscopic calculation is plotted versus porosity, considering $R_g = 0.05$ mm, initial $R_b = 0.02$ mm, and the two configurations with grains in direct contact ($d=0$) and not in direct contact ($d=0.01$ mm). In Figure 5b the corresponding curves obtained with eqn. (14) are shown, at several values of inter-granular distance $d$. Also in this case the curves obtained with the macroscopic relationship can adequately reproduce the microscopic trend, by setting parameters $\beta = 2$ and $\gamma = 0.67$.

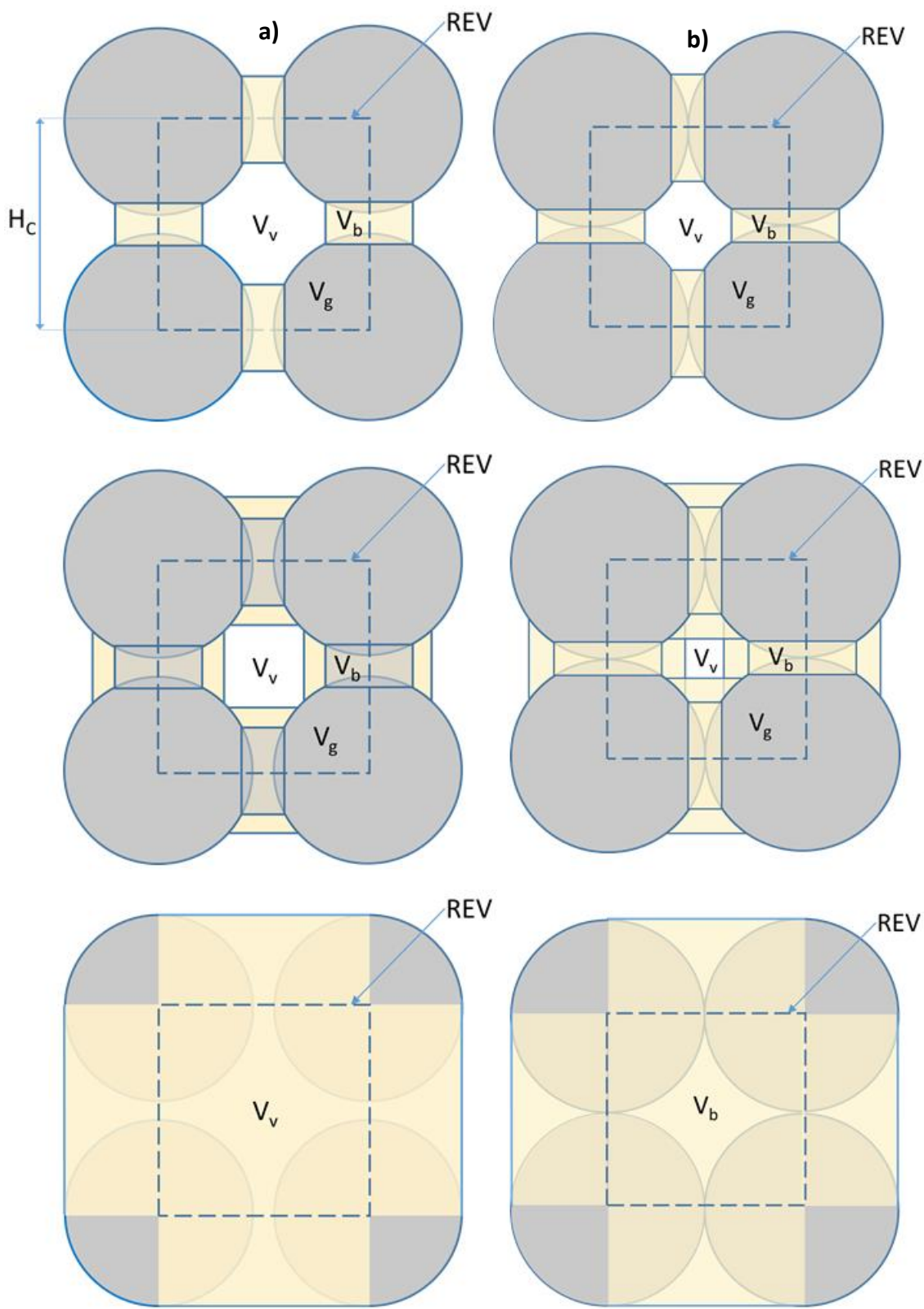

*Figure 3. 2D Schematic of the reference microscopic geometry for microscale calculations for the two ideal cases of (a) grains not in direct contact (d≠0) and (b) grains in direct contact (d=0). From top to bottom, the different figures represent different deposition phases, progressing from a large porosity configuration to a lower porosity configuration, ending with the complete filling of voids at the limit of zero porosity.*

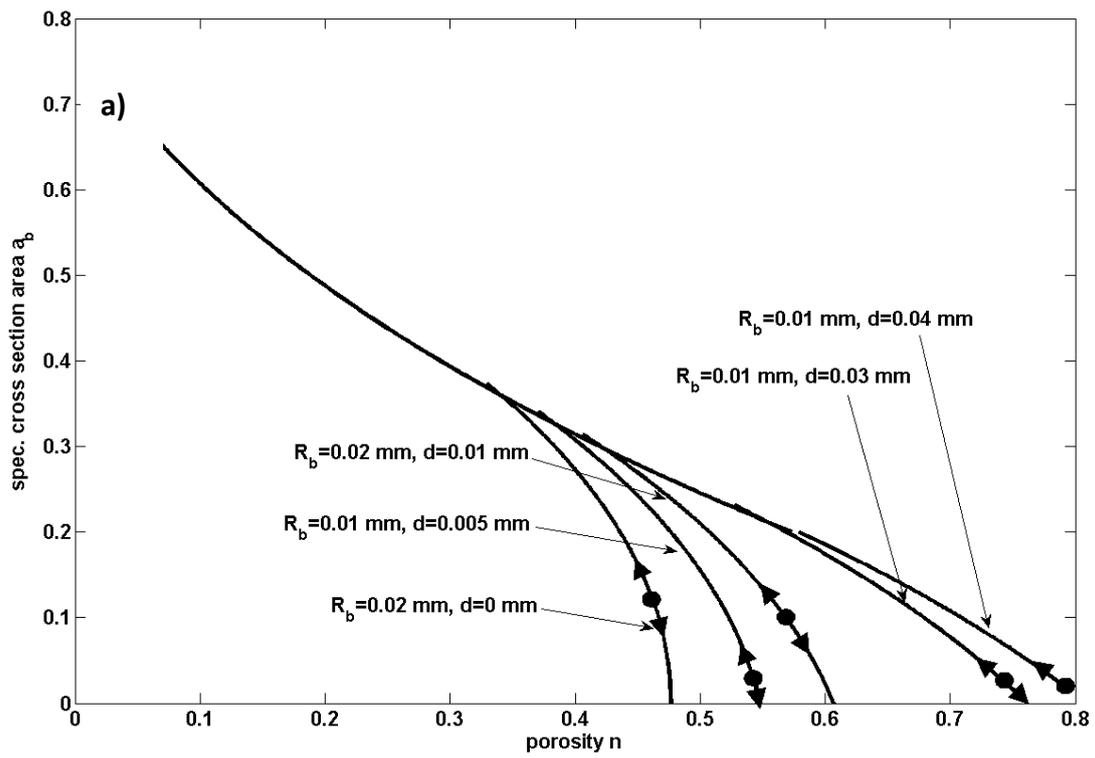
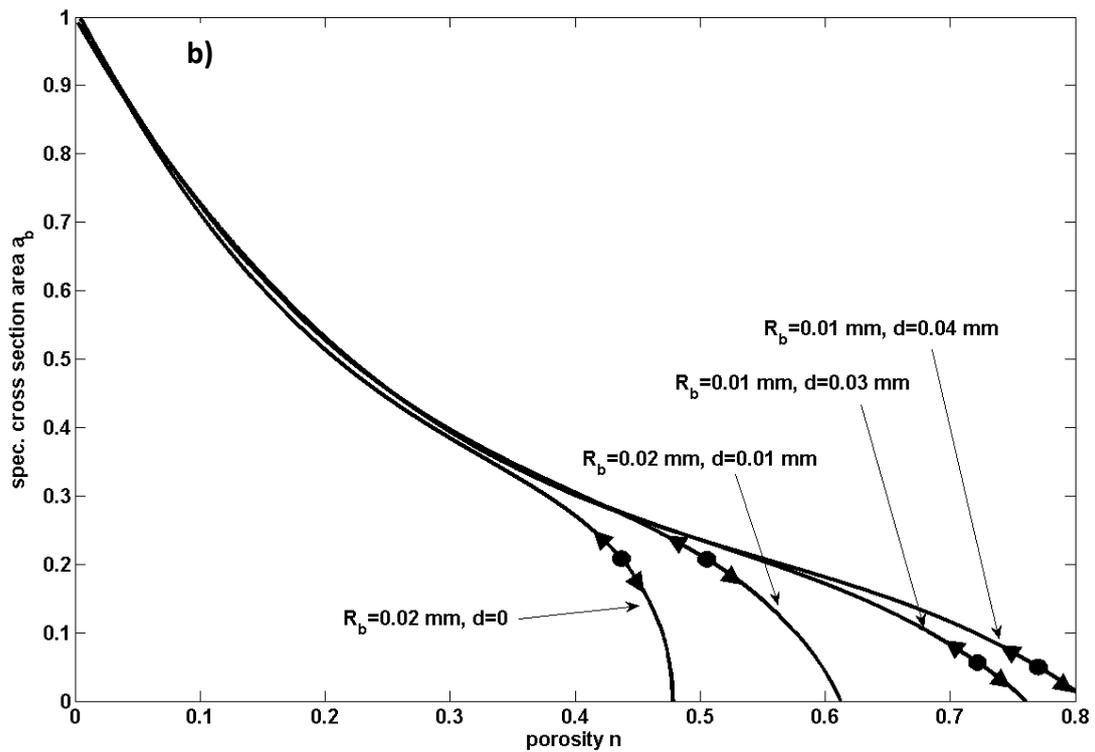

*Figure 4. Specific cross section area of cementing bonds versus porosity during dissolution/deposition (the dots mark the initial porosity). a) microscopic calculation, b) macroscopic calculation with model equation ( 16).*

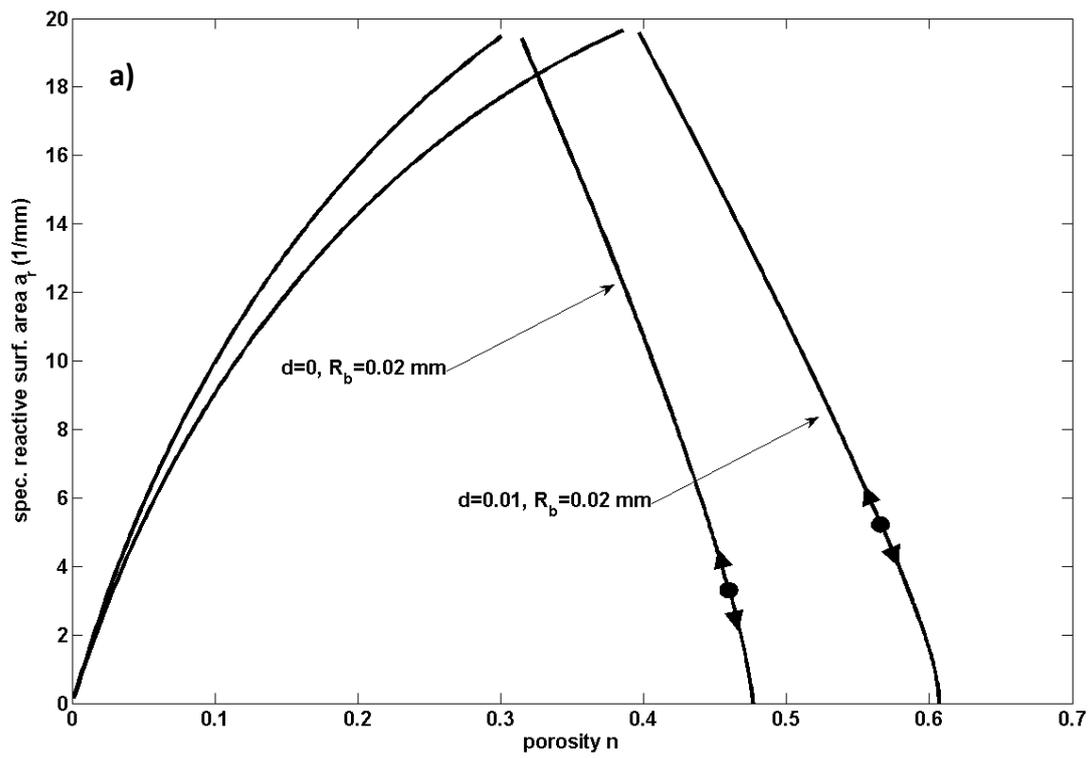

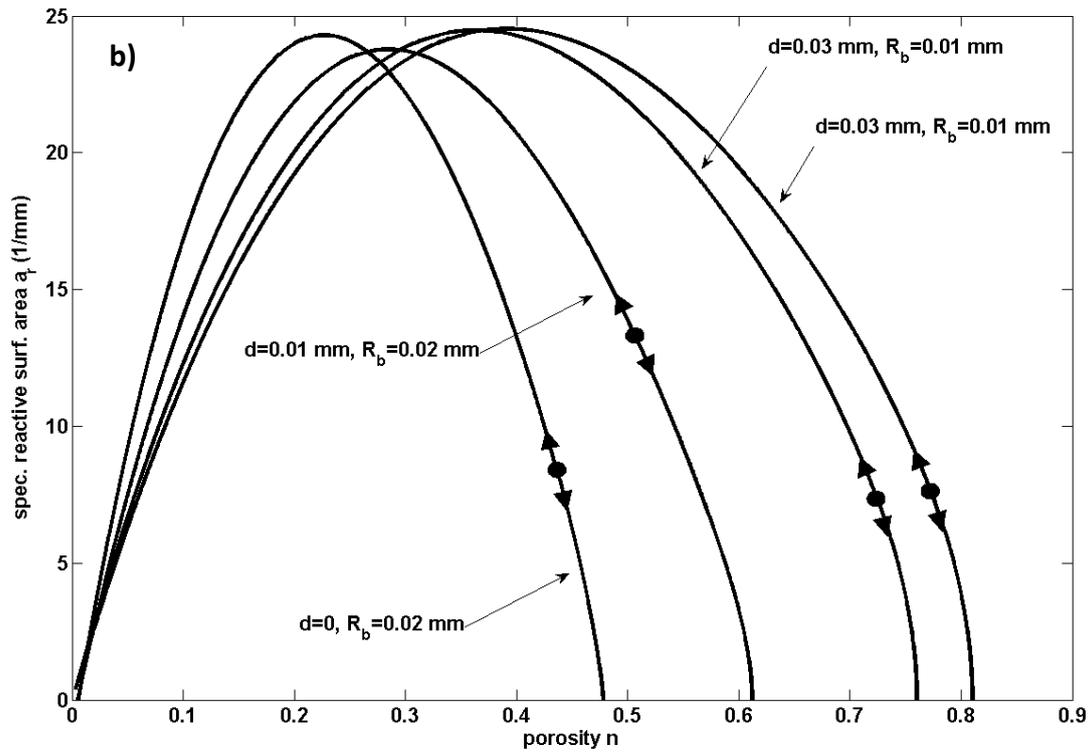

*Figure 5. Specific reactive surface area versus porosity during dissolution/deposition* (the dots mark the initial porosity). *a) microscopic calculation, b) macroscopic calculation with model equation* (14).

### 3. Model validation

The model described in Section 2 was numerically integrated, resorting to an automatic differentiation technique and automatic code generation system [50], [51] through a fully implicit, backward Euler integration scheme. The model capabilities were then tested by reproducing a series of chemo-mechanical experiments available in the literature.

The model parameters can be grouped into four categories, describing: (i) the material's uncemented behavior ($\lambda$, $\phi_{cv}$, $E_g$, $\nu_g$), (ii) the mechanical properties of the cementing material ($\sigma_{rc}$ and $\sigma_{rt}$) and of completely cemented soil ($E_b$), (iii) the mechanically-induced destructuration ($A$, $k_1$) and chemical healing ($k_2$), and (iv) the macroscopic evolution of specific reactive surface area and the specific cross section area of cementing bonds ($\beta$, $\gamma$, $\delta$, $\theta$, $a_{r0}$).

In addition to selecting the above parameters, initial values of variables describing the microscopic bond geometry (see Table 1) must be defined, possibly by examining thin sections and/or SEM images of the microstructure of the cemented materials. Useful information can be also obtained from chemical analyses, such as quantifying the sample solid mass variation at different degrees of chemical interaction. Parameters describing the mechanical properties of the (intact) cementing material may be selected from the rock mechanics literature, while those associated with material's uncemented behavior should be obtained from bespoke testing of the material in a fully destructured state, wherever possible. The interested reader is referred to [1] (see Section 5 therein) for a thorough discussion on the model parameter selection procedure.

### 3.1 Simulation of triaxial tests on artificially cemented sand

The model was first employed to reproduce chemo-mechanical loading paths in microbially cemented sands, thus simulating the strengthening and stiffening effect that artificially induced cement deposition has on granular soils. Reference was made in particular to an extensive set of experiments reported by Montoya and DeJong [27]. The authors performed a series of triaxial tests on Ottawa sand after subjecting it to different degrees of cementation, by means of ureolytic-driven calcite precipitation. After cementation completion, specimens were flushed with deaired water, back-pressured and then sheared in either undrained or drained conditions in a triaxial apparatus. The authors also performed measurements of shear wave velocity $V_s$ throughout shearing, to monitor small stiffness changes due to cementation's mechanical degradation.

Simulations were performed aimed at reproducing the experimental paths reported in [27] for drained triaxial testing of both untreated (uncemented) and microbially treated sand to a moderate cementation level, corresponding to an introduced amount of calcite of 0.6% by mass. The parameter values adopted for this set of simulations on calcarenite are shown in Table 2 (set #1a for treated sand and set #1b for untreated sand). In particular, the initial porosity was deduced from the initial void ratio estimated by [27] and the average grain radius was obtained from the measured D$_{50}$ of Ottawa sand [27]. Grains were described as having round shape, thus qualitatively being particularly consistent with our assumptions for the microscopic geometry. Among the different experimental datasets available in [27], drained triaxial compression experiments were selected, as these represent a more basic situation compared to their undrained counterpart, and can be considered a more didactic example of the key chemo-mechanical features of the model.

In Figure 6 the simulated deviatoric stress vs axial strain during drained triaxial compression is shown for both the untreated and the treated material, along with the corresponding experimental data. A higher peak strength and an overall more brittle behavior can be observed for the cemented material compared to the untreated material, while at large strain both materials tend to converge to the same strength value. These characteristic aspects are well captured by the model, both qualitatively and quantitatively. In Figure 7 the simulated and experimentally measured evolution of volumetric strain with axial strain is shown for both the untreated and the treated sand. Consistently with the more brittle character highlighted in Figure 6, the cemented sand exhibits a more markedly dilative behavior. Trends are correctly captured by the simulation, despite a tendency to slightly overpredict the initial contractive material behavior. Figure 8 shows simulations of shear wave velocity (calculated using standard elasticity relationships starting from elastic moduli, in turn obtained upon deriving the elastic free energy described in section 2.1.1) vs axial strain for both materials, that adequately reproduce the corresponding measured values.

A particularly good agreement between simulated and measured experimental trends is observed in Figure 6. In Figure 7 and Figure 8 the experimental data are reproduced somewhat less accurately. While the elasto-plastic evolution of volumetric strain, with particular reference to the untreated material, is adequately captured, the initial elastic volumetric straining appears over-predicted. This is mainly due to our model assumptions, that were made for the sake of simplicity and generality, e.g.: the adoption of an associated flow rule, the fact that we neglect elastic anisotropy and the increase of elastic stiffness due to stress level increase, as well as the bulk stiffness increase due to soil densification (this is also reflected by the unchanged shear wave velocity with increasing specimen compression, in the simulation of Figure 8, for untreated soil).

To gain further insight into the relationships between the simulated evolution of bonds and the macroscopic mechanical response, in Figure 9 the ratio $N_{ba}/N_b$ is plotted against axial strain for the above discussed simulations, calibrated on Montoya and DeJong's [27] experiments. It can be observed that during initial elastic loading no bond rupture occurs, while upon elasto-plastic shearing, mechanical destructuration occurs as the number of active bonds steadily decreases with increasing axial strain. However, the rate of destructuration decreases with shearing, to asymptotically tend to a tableau, at very large strain, representing fully destructured material.

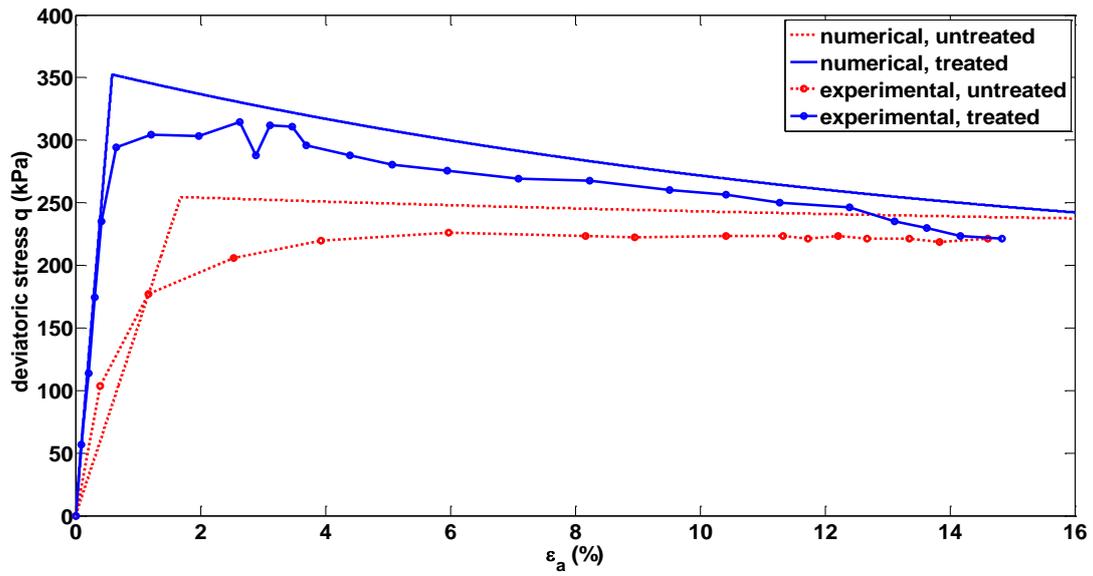

*Figure 6. Simulations and experimental data [27] of deviatoric stress vs axial strain during a drained triaxial test on uncemented and microbially cemented sand.*

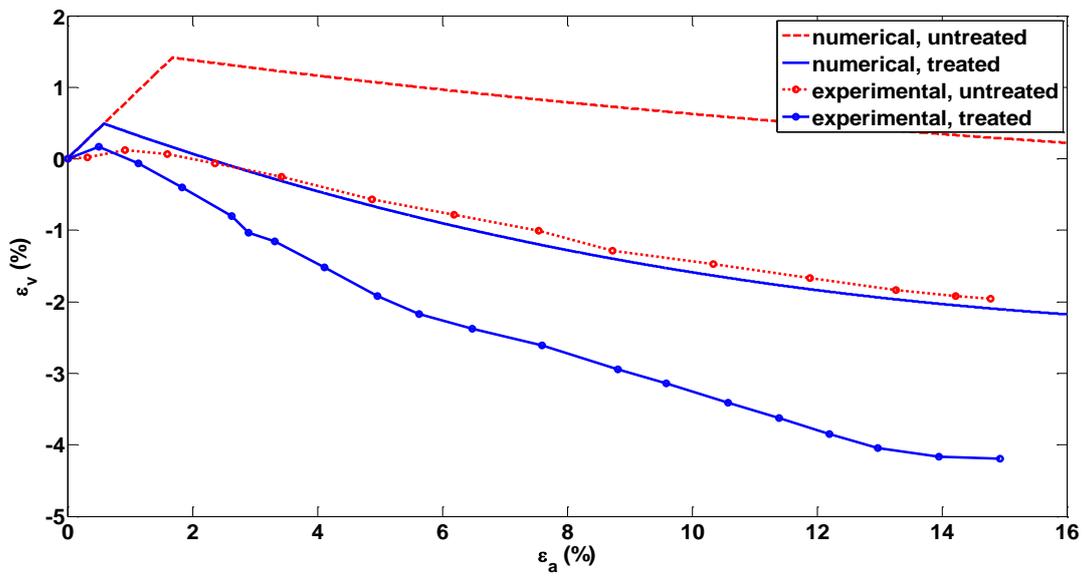

*Figure 7. Simulations and experimental data [27] of volumetric strain vs axial strain during a drained triaxial test on uncemented and microbially cemented sand.*

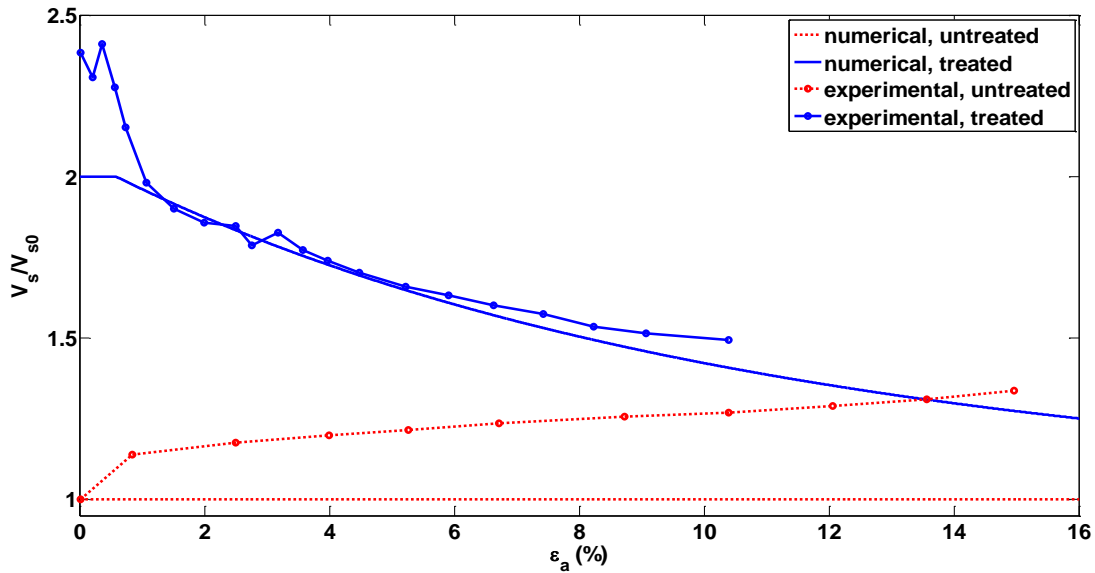

*Figure 8. Simulations and experimental data [27] of the evolution of normalized shear wave velocity with axial strain during a drained triaxial test on uncemented and microbially cemented sand.*

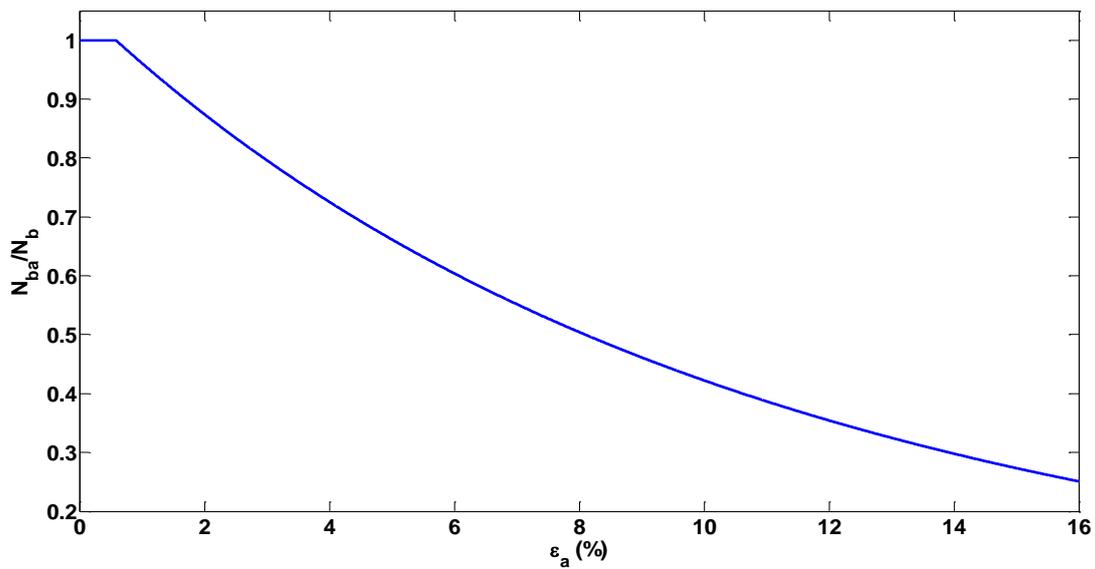

*Figure 9. Simulated evolution of the normalized number of active bonds with axial strain during drained triaxial shearing, for microbially cemented sand.*

### 3.2 Simulation of oedometer tests on artificially cemented sand

Another significant set of published chemo-mechanical experimental data come from Castellanza [28] and Castellanza and Nova's [29] special oedometer tests on artificially cemented sand. In this case, artificial cementation was achieved upon treating samples of coarse silica sand with lime and distilled water, thus obtaining a cemented material in a controlled manner, upon lime hydration. Specimens directly prepared within an instrumented oedometer thin ring (enabling measurement of the radial stress) were subjected to a chemical debonding test, consisting of first applying a mechanical vertical load and subsequently imposing a constant acid flow through the sample, at constant axial load. To simulate these experiments, the needed parameter values were obtained whenever possible from [28] and [29], and are shown in Table 2 (set #2).

In Figure 10 the simulation and corresponding experimental data are shown in terms of radial stress versus weathering index $\xi$ (amount of dissolved cement with respect to the total initial cement) during the chemical debonding phase. Radial stress keeps initially constant but soon starts to increase with a steep slope, to then switch to slightly increasing with a much gentler slope in the range $\xi = 0.5 - 1$. Simulations can be observed to consistently reproduce the measured data. To explore the impact of varying the average intergranular distance, simulations were performed both at $d = 0$ (corresponding to cement being deposited around grain-grain contacts only) and at $d = 0.01$ mm (indicating the presence of cement also bridging the gap between parts of grains not originally in contact). The latter setting corresponds to a better accuracy of reproduction of experimental data. Further, in Figure 11 the numerical and experimental evolution of axial strain with weathering index $\xi$ during the chemical debonding phase is shown. It can be noticed that initially, during the abrupt radial stress increase observed in Figure 10, the simulated axial strain increase in Figure 11 is very small (although not zero), while as radial stress stabilizes (roughly at $\xi \approx 0.5$), large plastic axial strains are triggered. The relative better accuracy of data reproduction of simulations with $d = 0.01$ mm also in Figure 11 confirms the aptness of this parameter choice for the soil at hand.

Based on the above results, it can be deduced that the present framework can be applied to adequately reproducing the chemo-mechanical behavior of lime-treated soils, even though this soil improvement procedure significantly differs from that of microbially cemented sands. Moreover, while in Section 3.1 simulations reproduce soil mechanical loading as a separate (i.e. subsequent) step from chemical treatment, in this section's simulations chemical and mechanical loading occur at the same time. This highlights the flexibility of the model to deal with different practical situations, as well as different kinds of bonded materials.

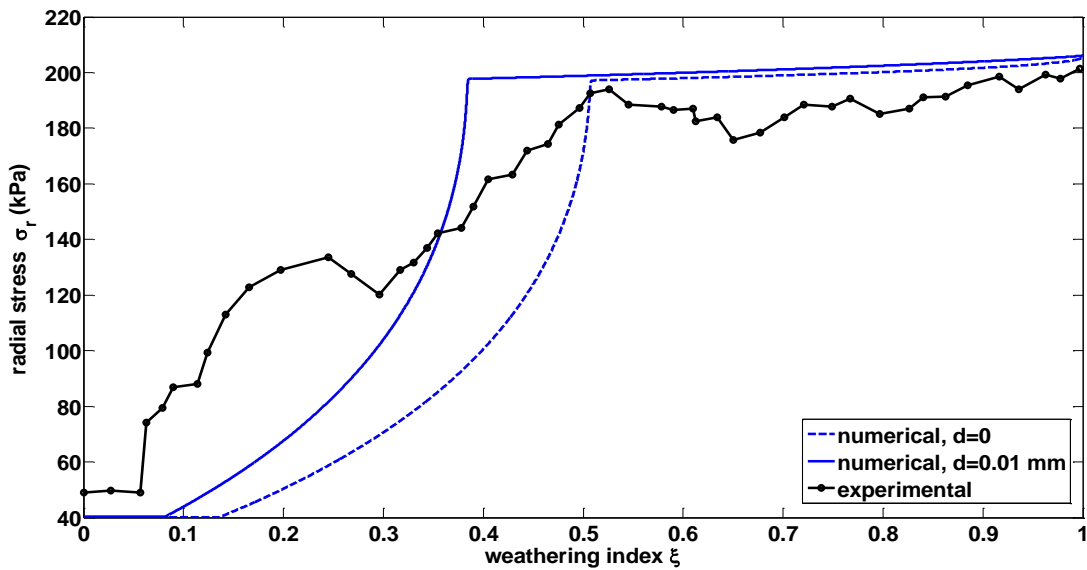

*Figure 10. Simulations and experimental data [28] of the evolution of radial stress with acid weathering in oedometer conditions on an artificially cemented sand.*

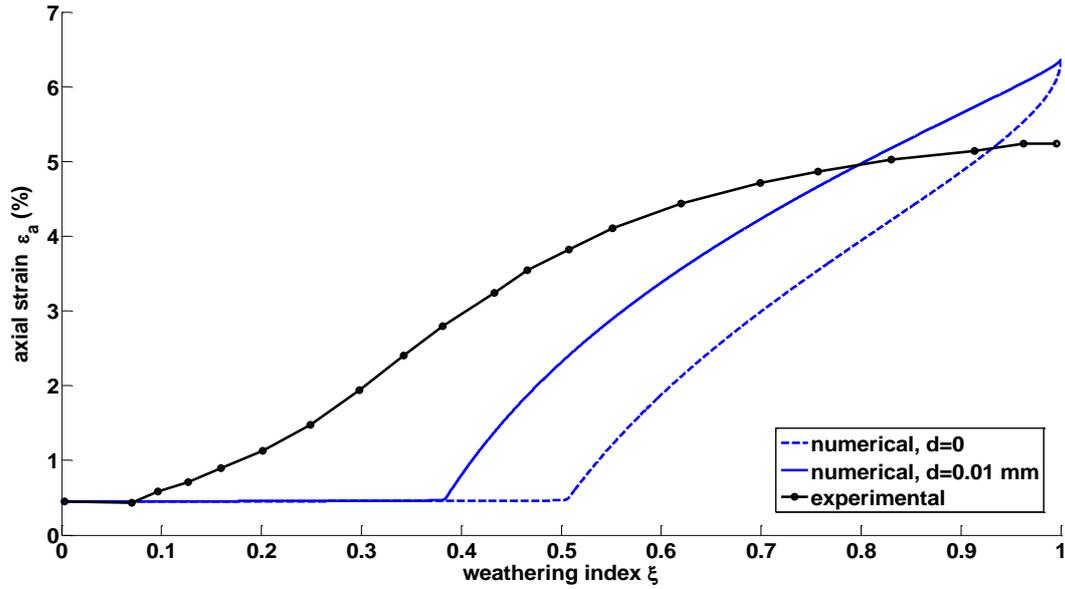

*Figure 11. Simulations and experimental data [28] of the evolution of axial strain with acid weathering in oedometer conditions on an artificially cemented sand.*

### 3.3 Simulation of uniaxial compression and oedometer tests on natural bonded soils

As an additional model validation, the experimental results of Ning et al. [30] on a natural soft rock are considered. The authors performed chemo-mechanical tests on a naturally calcite-cemented 'arkosic' sandstone, by subjecting different rock samples to weathering upon immersing them into acid solutions of different pH values, during different weathering periods, and subsequently testing them mechanically in uniaxial compression conditions. The effects of both acidity of the weathering agent and of total exposure time to weathering were thus explored in terms of material strength and deformability evolution. Parameter settings for this dataset are reported in Table 2 (set #3).

In Figure 12a the simulated deviatoric (axial) stress versus axial strain is plotted at different degrees of weathering (in the range $\xi = 0 - 0.9$ ), showing a marked decrease of both elastic stiffness and of peak strength with increasing $\xi$. The material can be observed to undergo a transition between a markedly brittle behavior and a more ductile one. Figure 12b shows Ning et al.'s [30] experimental results on the arkosic sandstone for comparison, tested after different weathering times in an acid solution with pH=5. It emerges from Figure 12 that experimental curves exhibit a more brittle behavior than simulated ones, however this is due to the typical inability of the adopted constitutive model to accurately reproduce the effects of strain localization. In fact, despite the underlying model assumption of a homogeneously deforming sample, the rock specimens at hand, tested under uniaxial compression conditions, are most likely to have developed shear bands.

In Figure 13 the simulated and experimentally measured variation of secant elastic modulus $E$ with weathering time is reported, while Figure 14 shows the simulated and experimentally measured variation of material's UCS with weathering time. It should be observed that since in [30] experimental results are expressed in terms of time of weathering (days), an additional calculation was performed to relate absolute time and the weathering index in our model, to be able to express the simulated evolution of $E$ and UCS as a function of weathering time (instead of $\xi$), as detailed in Appendix D. Also in this case, the model demonstrates the ability to consistently reproduce different types of chemo-mechanical experimental paths.

Finally, to test the model capabilities in the case of purely mechanical loading of bonded geomaterials, an experimental loading path obtained in oedometer conditions on a natural soft rock by [29] was numerically reproduced. Relevant parameter settings, determined after [29] (see [1] for further details), are reported in Table 2 (set #4). Figure 15 shows a direct comparison of the simulated and measured data in terms of axial strain vs axial stress. Good agreement between the two datasets is apparent, demonstrating that the proposed model is able to capture the typical sudden axial collapse, occurring when the bulk of mechanically induced destructuration takes place.

*Table 2. Parameter values adopted for model validation simulations.*

| Parameter | Symbol | Set #1a | Set #1b | Set #2 | Set #3 | Set #4 | Units |
|---|---|---|---|---|---|---|---|
| Initial porosity | $n_0$ | 0.73 | 0.73 | 0.475 | 0.395 | 0.56 | |
| Elastic modulus of uncemented solid skeleton | $E_g$ | 15 | 15 | 1 | 5 | 1 | MPa |
| Elastic modulus of cement material | $E_b$ | 7.7 | 7.7 | 4 | 53 | 150 | GPa |
| Poisson ratio of solid skeleton and bonds | $\nu_g, \nu_b$ | 0.08 | 0.08 | 0.08 | 0.08 | 0.053 | |
| Critical state parameter | $M_{cv}$ | 1.20 | 1.20 | 1.85 | 1.85 | 1.51 | |
| Initial mean radius of grains | $R_g$ | 0.10 | 0.10 | 0.05 | 0.2 | 0.05 | mm |
| Initial mean radius of bond cross-sections | $R_b$ | 0.021 | 0.021 | 0.02 | 0.073 | 0.009 | mm |
| Mean bond length | $d$ | 0.0 | 0.0 | 0.0 | 0.02 | 0.008 | mm |
| Destructuration plastic strain parameter | $A$ | 0.5 | 0.5 | 0.8 | 0.5 | 0.8 | |
| Mechanical destructuration rate parameter | $k_1$ | 10 | 10 | 0.8 | 70 | 0.8 | |
| Chemical healing rate parameter | $k_2$ | 0 | 0 | 0 | 0 | 0 | m³/kg |
| Slope of normal consolidation line | $\lambda$ | 0.15 | 0.15 | 0.03 | 0.03 | 0.05 | |
| Number of grains | $N_g$ | 6.4x10¹⁰ | 6.4x10¹⁰ | 1.0x10¹² | 1.8x10¹⁰ | 8.0x10¹¹ | m⁻³ |
| Number of bonds | $N_b$ | $=8 \times N_g$ | $=8 \times N_g$ | $=8 \times N_g$ | $=8 \times N_g$ | $=11 \times N_g$ | m⁻³ |
| Initial number of active bonds | $N_{ba}$ | $=N_b$ | $=10^5 \times N_b$ | $=N_b$ | $=N_b$ | $=N_b$ | m⁻³ |
| Tensile strength of cement material | $\sigma_{rt}$ | 0.15 | 0.15 | 0.1 | 43 | 8 | Mpa |
| Compressive strength of cement material | $\sigma_{rc}$ | 3 | 3 | 2 | 215 | 90 | Mpa |
| Uncemented preconsolidation stress | $p_c$ | 420 | 420 | 50 | 100 | 300 | kPa |
| Constitutive parameter | $\beta$ | 2 | 2 | 2 | 2 | 2 | |
| Constitutive parameter | $\gamma$ | 0.67 | 0.67 | 0.67 | 0.67 | 0.67 | |
| Constitutive parameter | $\delta$ | 1.5 | 1.5 | 0.5 | 0.5 | 0.5 | |
| Constitutive parameter | $\theta$ | 0.5 | 0.5 | 0.5 | 0.5 | 3.5 | |
| Constitutive parameter | $a_{r0}$ | 1 | 1 | 1 | 1 | 1 | 1/mm |

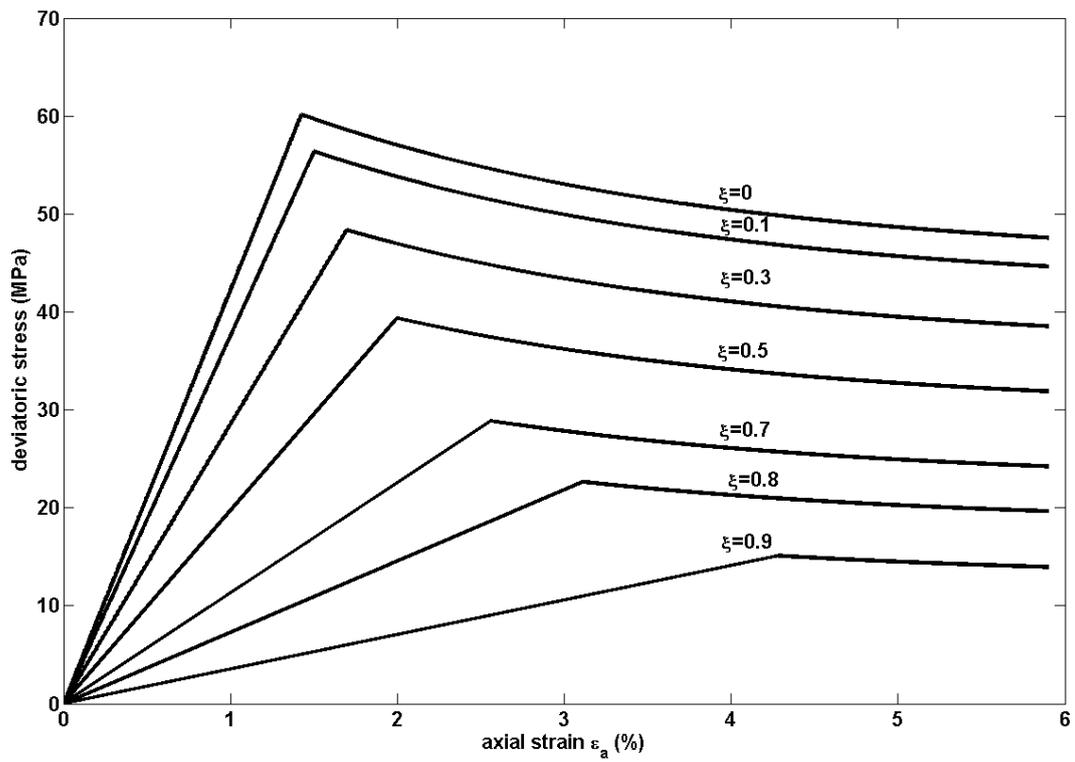

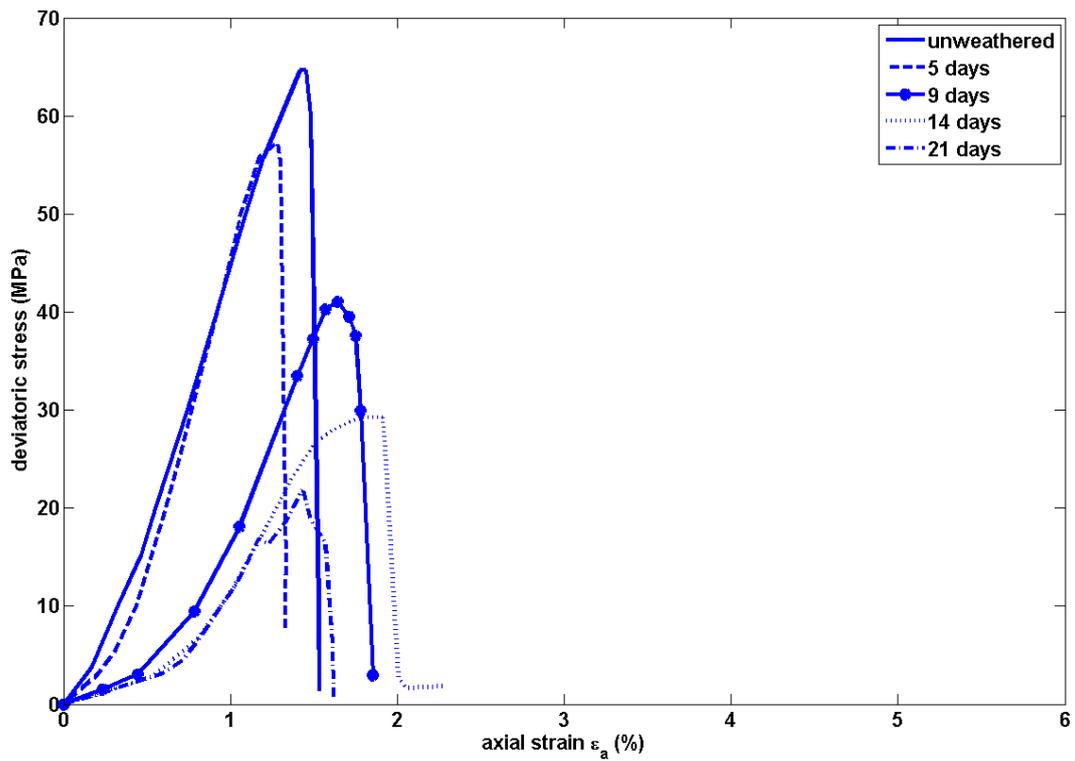

*Figure 12. (a) Simulations and (b) experimental data [30] of deviatoric stress vs axial strain during uniaxial compression tests on sandstone after subjecting samples to acid weathering for different periods.*

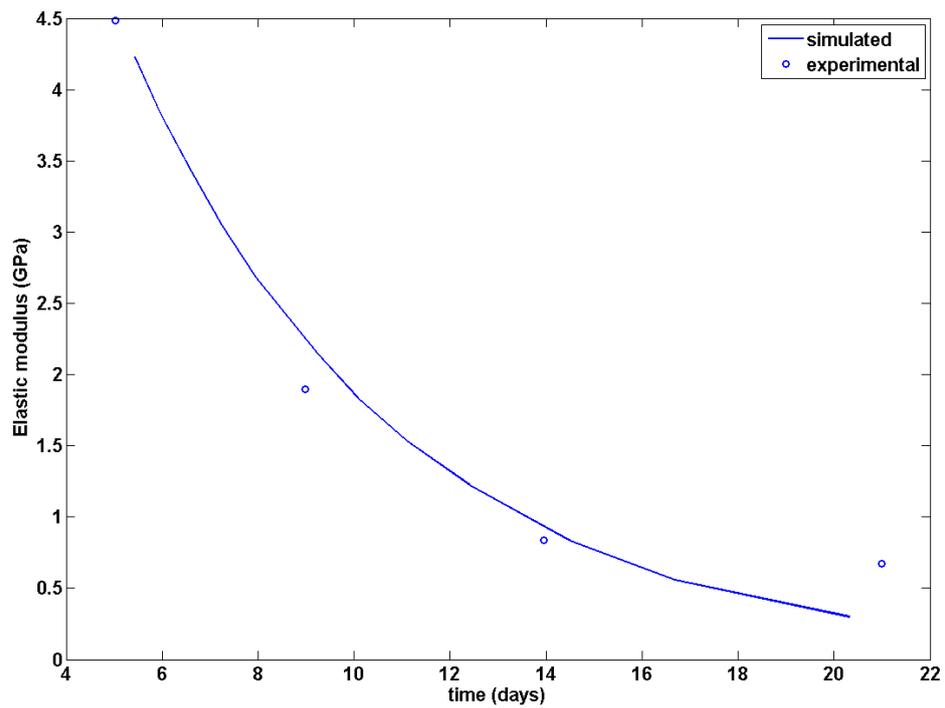

*Figure 13. Comparison of simulations and experimental data [30] in terms of elastic modulus versus weathering time. The experimental data are taken from Ning et al.'s [30] uniaxial compression tests on arkosic sandstone.*

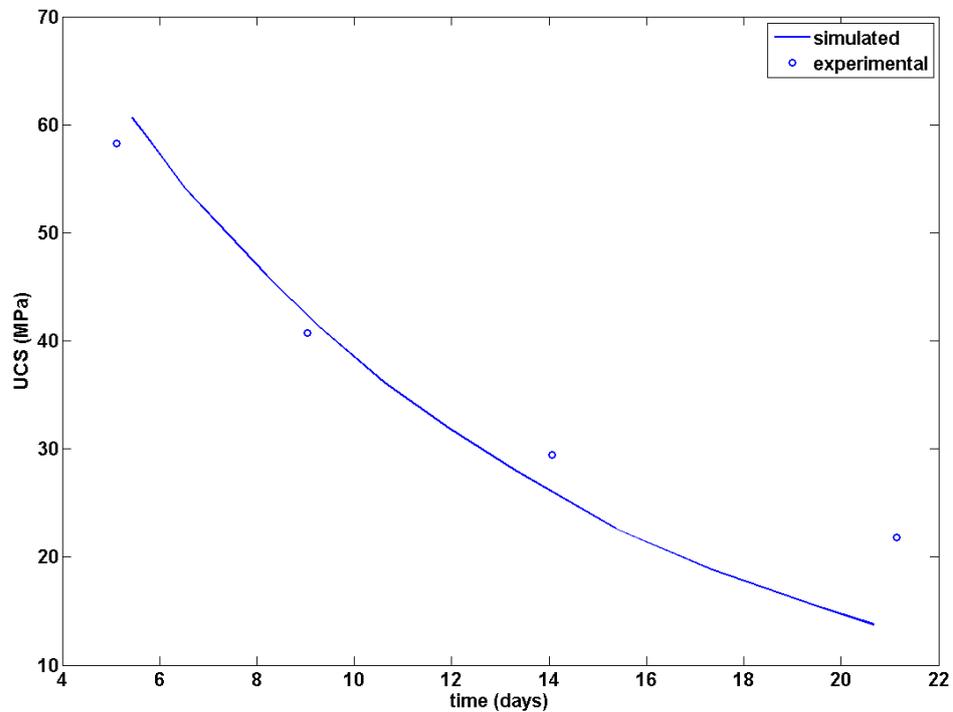

*Figure 14. Comparison of simulations and experimental data [30] in terms of UCS versus weathering time. The experimental data are taken from Ning et al.'s [30] uniaxial compression tests on arkosic sandstone.*

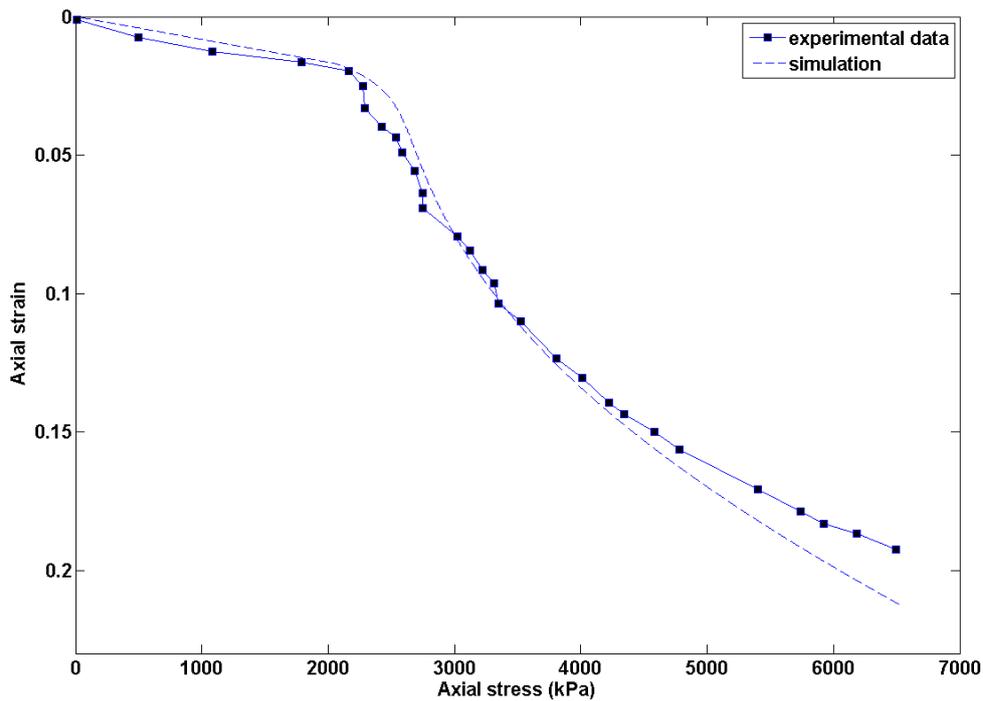

*Figure 15. Comparison of simulations and experimental data [29] in terms of axial strain versus axial stress in a purely mechanical oedometer test on unweathered soft rock.*

### 4. Conclusions

A dual scale chemo-mechanical constitutive model was presented, building up on the framework introduced by [1], able to reproduce the key features of the behavior of naturally or artificially cemented geomaterials subjected to both mechanical loading and cement deposition/dissolution due to chemical action. At variance with the bulk of existing literature, this model is specialized in representing cemented materials with nonreactive grains and reactive bonds, such as artificially (e.g., microbially) cemented silica sand and natural sandstone with carbonate cementing bonds.

A contribution of the modelling framework presented here, at variance with most existing chemo-mechanical models, is the capability to avoid unrealistic situations, such as further cement deposition at no available porous space, or further dissolution at no cement material left. This is achieved thanks to the introduction of two key cross-scale functions, namely the specific cross-sectional area and the specific reactive surface area of bonds.

Macroscopic plasticity is modeled through a modification of the standard critical state framework, while reversible behavior is described in the frame of hyperelasticity, allowing for the occurrence of chemo-elasto-plastic coupling. The effect of cement deposition on elasticity and the asymptotic behavior of specific reactive surface area close to null porosity have been examined particularly in depth in dedicated appendices.

Overall, the model allows keeping track of the evolution of the all-important reactive surface area, cross-sectional area and the number of mechanically active bonds along with dissolution/deposition, to reproduce the macroscopic chemo-mechanical behavior of cemented materials. Moreover, this framework allows deducing the porous medium unit weight variations consistently with porosity changes and mass dissolution/precipitation. The consistent simulation of a set of different experimental chemo-mechanical loading paths (involving both increased cementation and cement dissolution/destructuration) carried out on different (both naturally and artificially) cemented materials with only reactive bonds, demonstrates the reliability of this approach.

Although some simplifying assumptions were made, such as assuming linear elasticity, considering the bonds as mono-mineral and assuming an isotropic bond distribution, the proposed model could be promptly extended to more complex materials, such as those characterized by the presence of more than one family or anisotropic patterns of bonds, with different chemo-mechanical properties. It should be also remarked that modelling osmotic suction effects, typically occurring due to the exposure of fine-grained partially saturated soils to certain chemical solutions, is outside the scope of this work. Future developments will involve adopting more complex constitutive assumptions to deal with an extended notion of hydro-chemo-mechanical fatigue due to cyclic loading [50] superimposed on progressive (including geochemical) weakening [53] in slope instability, in the spirit of [54].

In conclusion, the presented model constitutes a step ahead towards the understanding of the behavior of bonded geomaterials subjected to both mechanical and chemical loading, and allows reproducing complex experimental chemo-mechanical paths in a reliable manner, yet retaining a relatively simple and didactic character that make it suitable to serve as a basis for further developments.

**Acknowledgements**

The first author would like to thank the Ministry for Education, University and Research (Ministero dell'Istruzione, dell' Università e della Ricerca - MIUR) for the partial financial support of this work under grant PRIN 2015: "Innovative Monitoring and Design Strategies for Sustainable Landslide Risk Mitigation".

The second author gratefully acknowledges financial support from the EU research project ERC-2013-ADG-340561-INSTABILITIES.

The third author gratefully acknowledges financial support from the "Visiting Professor 2013" program of the University of Trento.

**Appendix A. Effects of cement deposition on elasticity**

To further illustrate the implications of cement deposition in the material elastic behavior, let us consider a bonded porous medium in which the initial mean specific cross section of active cement bonds is $a_{b0}$ (that is the result of the previous chemical and mechanical history). If we set, for simplicity of presentation, $\alpha = 1$ in the expression of $\tilde{\omega}$ featuring in equation (2) (although the following rationale holds for any value of $\alpha$), then $\tilde{\omega} = a_b$, and the relevant initial elastic free energy (say at time $t_0$, hereafter called 'stage 0') is given by

$$\varphi(\boldsymbol{\varepsilon}_e, \boldsymbol{\varepsilon}_p, m_b) = (1 - a_{b0}) \, \varphi_g(\boldsymbol{\varepsilon}_e) + a_{b0} \, \varphi_b(\boldsymbol{\varepsilon}_e). \tag{A1}$$

Next, let us assume that in the time history of interest for our simulations, at time $t_1$ ('stage 1') cement deposition at constant (elastic) strain $\boldsymbol{\varepsilon}_{e,1}$ (under the stress state $\boldsymbol{\sigma}_1$) occurs, which leads to an increase of the mean cross section of active cement bonds from $a_{b0}$ to $a_{b1}$. After cement deposition, the free energy is assumed to be expressed by

$$\varphi\left(\boldsymbol{\varepsilon}_e, \boldsymbol{\varepsilon}_p, m_b\right) = (1-a_{b1})\ \varphi_g\left(\boldsymbol{\varepsilon}_e\right) + a_{b0}\ \varphi_b\left(\boldsymbol{\varepsilon}_e\right) + \left(a_{b1} - a_{b0}\right)\ \varphi_b\left(\boldsymbol{\varepsilon}_e - \boldsymbol{\varepsilon}_{e,1}\right). \tag{A2}$$

In this way, the newly deposited portion of cement bonds is initially unloaded and becomes active (i.e. loaded) only under a newly applied stress, that leads to an elastic strain change $\left(\boldsymbol{\varepsilon}_e - \boldsymbol{\varepsilon}_{e,1}\right)$. Thus, the elastic strain level existing during cement deposition, $\boldsymbol{\varepsilon}_{e,1}$ plays the role of an initial strain for the newly deposited portion of the cement bonds. It should be remarked that cement deposition increases soil stiffness.

Equivalently, if at a later time $t_2$, or 'stage 2' (at constant strain $\boldsymbol{\varepsilon}_{e,2}$, under the stress state $\boldsymbol{\sigma}_2$), a further cement deposition occurs, causing an increase of the specific mean cross section area of cementing bonds from $a_{b1}$ to $a_{b2}$, the final free energy becomes

$$\begin{aligned}\varphi\left(\boldsymbol{\varepsilon}_e, \boldsymbol{\varepsilon}_p, m_b\right) = (1-a_{b2})\ \varphi_g\left(\boldsymbol{\varepsilon}_e\right) + a_{b0}\ \varphi_b\left(\boldsymbol{\varepsilon}_e\right) + \left(a_{b1} - a_{b0}\right)\ \varphi_b\left(\boldsymbol{\varepsilon}_e - \boldsymbol{\varepsilon}_{e,1}\right) + \\ + \left(a_{b2} - a_{b1}\right)\ \varphi_b\left(\boldsymbol{\varepsilon}_e - \boldsymbol{\varepsilon}_{e,2}\right)\end{aligned} \tag{A3}$$

Thus we can easily generalize eqn. (A3) to a sum of *n*-terms or to a convolution integral (in the case for instance of cement deposition under smoothly changing applied strains or applied stresses).

The variables of interest can then be deduced from the free energy function in the standard way. As an example, let us assume that the free energy density of the uncemented grains $\varphi_g$ and of completely cemented soil $\varphi_b$ is the standard one of linear elasticity with Young moduli $E_g$ and $E_b$, and Poisson ratios, $\nu_g$ and $\nu_b$, respectively. Then, from eqns. (A1), (A2) and (A3), and deducing the stress $\boldsymbol{\sigma} = \partial \varphi / \partial \boldsymbol{\varepsilon}_e$, for uniaxial compression we would respectively obtain the following stress-strain relationships:

At 'stage 0' 
$$\sigma_{11} = (1-a_{b0})\ E_g \varepsilon_{11} + a_{b0}\ E_b \varepsilon_{11} \tag{A4}$$

At 'stage 1' 
$$\sigma_{11} = (1-a_{b1})\ E_g \varepsilon_{11} + a_{b0}\ E_b \varepsilon_{11} + \left(a_{b1} - a_{b0}\right)\ E_b \left(\varepsilon_{11} - \varepsilon_{e11,1}\right) \tag{A5}$$

At 'stage 2' 
$$\begin{aligned}\sigma_{11} = (1-a_{b2})\ E_g \varepsilon_{11} + a_{b0}\ E_b \varepsilon_{e11} + \left(a_{b1} - a_{b0}\right)\ E_b \left(\varepsilon_{e11} - \varepsilon_{e11,1}\right) + \\ + \left(a_{b2} - a_{b1}\right)\ E_b \left(\varepsilon_{e11} - \varepsilon_{e11,2}\right)\end{aligned} \tag{A6}$$

It is worth observing that, upon unloading from a compressive state after cementation, when $\varepsilon_{e11}$ becomes smaller than $\varepsilon_{e11,2}$ or $\varepsilon_{e11,1}$, then some terms of eqns. (5) and (6) become negative, i.e. provide a local traction.

**Appendix B. Hardening law**

To derive the hardening rule in a more general form that would be inclusive of material structural changes, primarily, mineral mass addition or removal, we still assume that the evolution of preconsolidation pressure $p_c$ depends on the evolution of voids in the material, but in a broader sense. As initially the contribution of mineral mass dissolution/deposition to the voids evolution is not considered, the isotropic normal consolidation line in a $e - \log p$ plot can be classically defined in a differential form as follows (e.g. see [55])

$$\delta e = -\lambda \frac{\delta p_c}{p_c} \tag{B1}$$

which is valid for $p_c = p$. In terms of total volumetric strain, since at small strains $\delta \varepsilon_v = \delta e / (1+e_0)$, the relationship can be rearranged into

$$\frac{\delta p_c}{p_c} = -\frac{(1+e_0)}{\lambda} \delta \varepsilon_v \tag{B2}$$

Next, upon including the contribution of dissolution/deposition to the void space variation, we must consider that dissolution/deposition leads to an increase/decrease of void space. Note that $p_c$ and $\lambda$ describe the behavior of the destructured soils, in which the broken grain bonds become part of the solid fraction affecting the grain size distribution. Thus, in general, during chemo-mechanical loading neither the slope of the isotropic compression line nor its intercept with the void ratio axis are bound to remain constant (e.g., see [56]). In particular, the slope $\lambda$ is expected to depend on the amount of broken grain bonds, namely $\lambda = \lambda(m_b)$. However, if we assume for the sake of simplicity that the subtraction/addition of solid phase does not change the grain size distribution and the shape of the particle surface, and that the different crushability of grains and broken bonds is not an issue, then the slope $\lambda$ can be assumed constant, thus

$$\frac{\delta p_c}{p_c} = \frac{1+e_0}{\lambda} \left( -\delta \varepsilon_v + \frac{\delta m_b}{\rho_s} \right) \quad \text{valid for } p_c = p \tag{B3}$$

where $\delta m_b$ is the increase of relative mass (positive for deposition, implying a decrease of void ratio) referred to the unit initial volume as in the definition of $\varepsilon_v$, and $\rho_s$ is the (constant) density of the solid grains and cementing material. It is worth emphasizing that eqn. (B3) holds true only for the stress states lying on the normal consolidation line, namely for $p_c = p$.

By exploiting the additive decomposition of strain increments, $\delta \varepsilon_v = \delta \varepsilon_v^e + \delta \varepsilon_v^p$, the elastic volumetric strain increment can be expressed in a mean stress-dependent form (like in the standard Cam Clay formulation) as

$$\delta \varepsilon_v^e = -\frac{\kappa}{1+e_0} \frac{\delta p}{p} \tag{B4}$$

representing the equation of the isotropic unloading/reloading line.

For a stress state lying on the isotropic normal consolidation line and for a destructured soil, which is subjected to a small virgin loading/elastic unloading cycle, the plastic volumetric strain increment can be obtained by substituting eqn. (B4) into eqn.(B3), thus obtaining the hardening relationship, expressed in a rate form

$$\frac{\dot{p}_c}{p_c} = (1+e_0)\left(-\frac{\dot{\varepsilon}_v^p}{\lambda-\kappa} + \frac{\dot{m}_b}{\lambda\rho_s}\right) \tag{B5}$$

where $\dot{p}_c$ is the rate of change of the preconsolidation pressure.

It is worth emphasizing that for $\dot{m}_b = 0$, the usual Cam clay hardening relationship is recovered from eqn. (B5), whereas for cement dissolution ($\dot{m} < 0$) in a cemented soil (with $N_{ba} > 0$) at constant isotropic mean stress $p$, $\dot{p}_c < 0$ as long as $p < p_c + p_{comp}$. Furthermore, in a continuing plastic process, when $p = p_c + p_{comp}$, plastic volumetric compression ($\dot{\varepsilon}_v^p < 0$) must occur thus $p_c$ increases, $\dot{p}_c > 0$, whereas $p_{comp}$ decreases due to the rupture of grain bonds. In any case, for the sake of consistency, $p = p_c + p_{comp}$ holds true (cf. Figure 1). As a result, cement dissolution induces plastic volumetric compression because of two mechanisms, due to (i) the rupture of cementation bonds (leading to a decrease of $p_{comp}$), and (ii) the dependence of $p_c$ on cement dissolution. Hence, in general, the combination of both mechanisms may lead to a dramatic volumetric compression because of bond rupture.

In this context, the consistency condition becomes

$$\dot{F} = \frac{\partial F}{\partial \sigma}\dot{\sigma} + \frac{\partial F}{\partial p_c}\dot{p}_c + \frac{\partial F}{\partial p_{comp}}\dot{p}_{comp} + \frac{\partial F}{\partial p_{tens}}\dot{p}_{tens} = 0 \tag{B6}$$

where $p_c = p_c(\varepsilon_v^p, m_b)$, $p_{comp} = p_{comp}(\varepsilon^p, m_b)$ and $p_{tens} = p_{tens}(\varepsilon^p, m_b)$.

If instead, for simplicity, we consider stress-independent elasticity, so that the bulk modulus $K$ is assumed constant (as is also assumed for the shear modulus $G$), by exploiting the additive decomposition of strain $\varepsilon_v = \varepsilon_v^e + \varepsilon_v^p$ and expressing the elastic volumetric strain as $\dot{\varepsilon}_v^e = -\dot{p}_c/K$, we can modify eqn. (B5) as

$$\frac{\dot{p}_c}{p_c} = -\frac{(1+e_0)}{\lambda}\left(\dot{\varepsilon}_v^p + \frac{\dot{p}_c}{K} - \frac{\dot{m}_b}{\rho_s}\right) \tag{B7}$$

Leading to

$$\dot{p}_c = \frac{\dfrac{\dot{m}_b}{\rho_s} - \dot{\varepsilon}_v^p}{\dfrac{\lambda}{p_c(1+e_0)} + \dfrac{1}{K}} . \tag{B8}$$

Moreover, we can assume that terms $1/K$ and $\dot{m}_b/\rho_s$ are negligibly small compared to the other terms in eqn. (B8), thus finally obtaining

$$\frac{\dot{p}_c}{p_c} = -\frac{1+e_0}{\lambda}\dot{\varepsilon}_v^p \tag{B9}$$

which leads upon integration to the hardening law (5).

**Appendix C. Asymptotic relationships**

To gain further insight on the asymptotic behavior of the specific reactive surface area close to null porosity ($n \to 0$), we can assume that in this range of porosities the porous space is described by a number of spherical voids per unit bulk volume $N_v$, with mean radius equal to $R_v$. As a result, the reactive surface area becomes equal to

$$a_r = N_v 4\pi R_v^2 \tag{C1}$$

and the void bulk volume becomes $v_v = N_v \frac{4}{3}\pi R_v^3$, which coincides with $n$. As a result, the ratio $a_r/v_v$ becomes

$$\frac{a_r}{n} = \frac{3}{R_v} \tag{C2}$$

Taking into account that from eqn. (C1) $\frac{\dot{a}_r}{a_r} = 2\frac{\dot{R}_v}{R_v}$ and that the void volume variation $\dot{V}_v$ is related to the mean radius variation and to the void surface through $\dot{V}_v = \dot{R}_v a_r$, then

$$\dot{R}_v = -\frac{\dot{m}_b}{a_r \rho_s} \quad \text{with} \quad \dot{n} = -\frac{\dot{m}_b}{\rho_s} \tag{C3}$$

where the volume strain of the solid skeleton has been neglected in eqn. (C3)b. From eqns. (1) and (2) we obtain

$$\frac{\dot{a}_r}{a_r} = \frac{2}{3}\frac{\dot{n}}{n} \tag{C4}$$

thus, close to the $n \to 0$ limit, the asymptotic behavior is obtained integrating (C4) as $a_r = a_{r0} n^{2/3}$, with $a_{r0}$ a constitutive parameter.

It is worth adding that if the porous space were approximated to a series of cylinders of constant section, then the asymptotic behavior for $n \to 0$ would have been $a_r = a_{r0} n$. As a result, we can more generally assume

$$a_r = a_{r0} n^\gamma \tag{C5}$$

where the exponent $\gamma$ is a constitutive parameter.

Note that eqn. (C5) depending on $n$ (instead of $\tilde{n}$) implies that a volumetric strain induces a variation of reactive surface area (according to eqn. (8)). Although this assumption could be questionable (because a volumetric compression does not necessarily imply a reduction of $a_r$ and might even lead to an increase of $a_r$, if micro-cracking occurs) eqn. (C5) has been assumed valid because for $n \to 0$ it prevents further mass deposition, thus avoiding the unphysical situation of mass deposition in a porous medium with null void volume.

**Appendix D. Degree of weathering**

The relationship between the degree of weathering $\xi$ and time can be obtained starting from the approximate relation

$$\xi(t) = 1 - \frac{m_b}{m_{b0}} = 1 - \frac{m_{b0} - \int_0^t v_r a_r d\tau}{m_{b0}} \tag{D1}$$

where $m_{b0}$ and $m_b = m_b(t)$ respectively the initial and current cement mass, $t$ is time, $v_r$ the velocity of chemical reaction (assumed constant) and $a_r = a_r(\xi)$ the reactive surface area of bonds (since $\xi = \xi(m_b)$, also $a_r = a_r(m_b)$). It can be deduced from (D1) that

$$\xi(t) = \frac{\int_0^t v_r a_r d\tau}{m_{b0}}. \tag{D2}$$

Upon differentiating and integrating the above, we obtain

$$\int_{\xi_0}^{\xi} \frac{1}{a_r(\xi)} d\xi = \int_{t_0}^{t} \frac{v_r}{m_{b0}} d\tau. \tag{D3}$$

The relationship between the reactive surface area of bonds and the degree of dissolution was obtained from our model simulations with parameter settings of Table 2 (Set #3), and was found by interpolation to be approximately linear, as follows:

$$a_r(\xi) = 3 - 2.7\xi \tag{D4}$$

Substituting (D4) into (D3) and developing the integrals, we obtain

$$\frac{1}{2.7}\left[\ln(3 - 2.7\xi_0) - \ln(3 - 2.7\xi)\right] = \frac{v_r}{m_{b0}}(t - t_0) \tag{D5}$$

Which becomes, considering $\xi_0 = 0$,

$$\ln(\xi) = \ln 3 - 2.7\frac{v_r}{m_{b0}}(t - t_0). \tag{D6}$$

The above can be solved for $t$ to finally yield

$$t(\xi) = t_0 - \frac{m_{b0}}{2.7 v_r} \ln\left(\frac{\xi}{3}\right). \tag{D7}$$

Equation (D7) has been used to convert degree of dissolution into time in the simulations presented in Section 3.2.

**References**

[1] DeJong, JT, Fritzges MB and Nusslein K. Microbially induced cementation to control sand response to undrained shear. J. of Geotechnical and Geoenvironmental Eng. ASCE 2006; 132: 1381-1392

[2] Cheng L, Cord-Ruwisch R, Shahin MA. Cementation of sand soil by microbially induced calcite precipitation at various degrees of saturation. Canadian Geotechnical Journal. 2013; 50(1):81-90.

[3] Al Qabany A, Soga K. Effect of chemical treatment used in MICP on engineering properties of cemented soils. Géotechnique. 2013; 63(4):331.

[4] Gomez MG, Martinez BC, DeJong JT, Hunt CE, deVlaming LA, Major DW, Dworatzek SM. Field-scale bio-cementation tests to improve sands. Proceedings of the Institution of Civil Engineers- Ground Improvement. 2015; 168(3):206-16.

[5] Venuleo S, Laloui L, Terzis, D, Hueckel T and Hassan M. Microbially induced calcite precipitation effect on soil thermal conductivity. Géotechnique letters 2016; 6 (1), 39-44.

[6] Terzaghi K. Mechanism of landslides; Geol. Soc. Am. Engineering Geology 1950; Berkey, pp.89-123.

[7] Zhao Y, Cui P, Hu LB, Hueckel T. Multi-scale chemo-mechanical analysis of the slip surface of landslides in the Three Gorges, China. Science China Technological Sciences 2011; 54: 1757-1765

[8] Moore, R., The Chemical and Mineralogical Controls Upon The Residual Strength Of Pure And Natural Clays, Geotechnique 1996; 41(1): 35-47

[9] Moore R, Brunsden D, Physico-chemical effects on the behaviour of a coastal mudslide, Geotechnique 1996; 46(2): 259-278

[10] Chigira M, Oyama T; Mechanism and effect of chemical weathering of sedimentary rocks, Engineering Geology 2000; 55: 3-14

[11] Ciantia MO and Hueckel T. Weathering of submerged stressed calcarenites: chemo-mechanical coupling mechanisms, Geotechnique 2013; 63 (9): 768-785.

[12] Hueckel T and Hu LB. Feedback mechanisms in chemo-mechanical multi-scale modeling of soil and sediment compaction, Computers and Geotechnics 2009; 36, 2009: 934-943

[13] Castellanza R, Nova R, Gerolymatou E. An attempt to predict the failure time of abandoned mine pillars, Rock Mechanics And Rock Engineering 2008; 41(3): 377-401 DOI: 10.1007/s00603-007-0142-y

[14] Andre L, Azaroual M, Menjoz A. Numerical Simulations of the Thermal Impact of Supercritical CO2 Injection on Chemical Reactivity in a Carbonate Saline Reservoir. Transport in porous media 2010; 82 (1):247-274

[15] Grgic D. Influence of CO2 on the long-term chemomechanical behavior of an oolitic limestone. Journal of Geophysical Research 2011; 116, B07201.

[16] Canal J, Delgado-Martin J, Barrientos V, Juncosa R, Rodriguez-Cedrùn B, Falcon-Suarez I. Effect of supercritical CO2 on the Corvio sandstone in a flow-thru triaxial experiment 2014, in Rock Engineering and Rock Mechanics: Structures in and on Rcok Masses, Alejano, Perucho, Olalla & Jimenez (Eds), Taylor & Francis, London.

[17] Hu LB and Hueckel T. "Coupled chemo-mechanics in evolving permeability in geomaterials", Computational Geomechanics (ComGeo II) 2011; edited by Pietruszczak and Pande, Cavtat-Dubrovnik, 599-608

[18] Matsukura Y and Hirose T. Five year measurements of rock tablet weathering on a forested hillslope in a humid temperate region. Engng Geol. 1999; 55: 69–76.

[19] Kohler SJ, Dufaud F, Oelkers EH. An experimental study of illite dissolution kinetics as a function of pH from 1.4 to 12.4 and temperature from 5 to 50°C. Geochim Cosmochim Acta 2002; 67(19): 3583–3594

[20] Hueckel T. Coupling of elastic and plastic deformations of bulk solids 1976; Meccanica 11(4): 227-235

[21] Maier G and Hueckel T. Nonassociated and coupled flow rules of elastoplasticity for rock-like materials, Int. J. Rock. Mech. Min. & Sci. Geomech. Abstr. 1979; 16: 77-92

[22] Hueckel T. Discretized Kinematic Hardening in Cyclic Degradation of Rocks and Soils, Engineering Fracture Mechanics 1985; 21(4): 923-945.

[23] Hueckel T and Tutumluer E. Modeling of elastic anisotropy due to one-dimensional plastic consolidation of clays, Computers and Geotechnics 1994; 16: 311-349

[24] Gajo A, Bigoni D. A model for stress and plastic strain induced nonlinear, hyperelastic anisotropy in soils. International Journal For Numerical And Analytical Methods In Geomechanics 2008; 32 (7), p. 833-861

[25] Gajo A. Hyperelastic modelling of small-strain stiffness anisotropy of cyclically loaded sand. International Journal For Numerical And Analytical Methods In Geomechanics 2010; 34(2), 111-134.

[26] Gajo A, Cecinato F, Hueckel T, A micro-scale inspired chemo-mechanical model of bonded geomaterials, International Journal of Rock Mechanics & Mining Sciences 2015; 80: 425–438.

[27] Montoya BM, DeJong JT. Stress-strain behavior of sands cemented by microbially induced calcite precipitation, 2015; Journal of Geotechnical and Geoenvironmental Engineering, 141(6).

[28] Castellanza R. Weathering effects on the mechanical behaviour of bonded geomaterials: an experimental, theoretical and numerical study 2002; PhD thesis, Milan Polytechnic University, Italy.

[29] Castellanza R and Nova R. Oedometric tests on artificially weathered carbonatic soft rocks, J. of Geotechnical and Geoenvironmental Eng. ASCE 2004; 130(7): 28 – 739

[30] Ning L, Yunming Z, Bo S, Gunter S. A chemical damage model of sandstone in acid solution.

International Journal of Rock Mechanics and Mining Sciences 2003; 40(2): 243–249.

[31] Wang W, Liu TG, Shao JF. Effects of acid solution on the mechanical behavior of sandstone. J. Mater. Civ. Eng. ASCE, 2015; DOI 10.1061/(ASCE)MT.1943-5533.0001317

[32] Le Guen Y, Renard F, Hellmann R, Brosse E, Collombet M, Tisserand D, Gratier JP. Enhanced deformation of limestone and sandstone in the presence of high $P_{CO2}$ fluids, Journal of geophysical research, 2007; 112, B05421.

[33] Hangx SJT, Spiers CJ, Peach CJ. Creep of simulated reservoir sands and coupled chemical-mechanical effects of CO2 injection, Journal of Geophysical Research 2010; 115, B09205.

[34] Venuleo S, Laloui L, Terzis D, Hueckel T. and Hassa, M. Microbially induced calcite precipitation effect on soil thermal conductivity, Géotechnique Letters 2016; 6(1): 39-44.

[35] Guo R, Hueckel T. Silica polymer bonding of stressed silica grains: An early growth of intergranular tensile strength. Geomechanics for Energy and the Environment 2015; 1:48-59.

[36] Michalowski RL, Wang Z, Park D. and Nadukuru SS. Static Fatigue or Maturing of Contacts in Silica Sand. In GeoShanghai International Conference. Springer, Singapore; 2018: 911-918.

[37] Terzis D, Bernier-Latmani R and Laloui L. Fabric characteristics and mechanical response of bio-improved sand to various treatment conditions 2016; Géotechnique Letters, 6(1): 50-57.

[38] Terzis D and Laloui L. 3-D micro-architecture and mechanical response of soil cemented via microbial-induced calcite precipitation 2018; Scientific reports, 8(1), p.1416.

[39] Gens A, Nova R. "Conceptual bases for a constitutive model for bonded soils and weak rocks", in proc. International Symposium on Geotechnical Engineering of Hard Soils - Soft Rocks 1993; Athens, Greece.

[40] Lagioia R, Nova R. An experimental and theoretical study of the behaviour of a calcarenite in triaxial compression. Geotechnique 1995; 45(4): 633–648

[41] Nova R, Castellanza R, Tamagnini C. A constitutive model for bonded geomaterials subject to mechanical and/or chemical degradation. Int J Num Anal Meth Geomech 2003; 27(9):705–732

[42] Tamagnini C, Ciantia MO. Plasticity with generalized hardening: constitutive modeling and computational aspects, Acta Geotechnica 2016; 11(3): 595-623

[43] Ciantia MO, Di Prisco C, Extension of plasticity theory to debonding, grain dissolution, and chemical damage of calcarenites, International Journal for Numerical and Analytical Methods in Geomechanics 2016; 40: 315-343

[44] Voigt W. Ueber die Beziehung zwischen den beiden Elasticitätsconstanten isotroper Körper, Annalen der Physik 1889; 274: 573–587

[45] Reuss A. Berechnung der Fließgrenze von Mischkristallen auf Grund der Plastizitätsbedingung für Einkristalle. ZAMM - Journal of Applied Mathematics and Mechanics / Zeitschrift für Angewandte Mathematik und Mechanik 1929; 9: 49–58.

[46] Nova R. "Modelling the weathering effects on the mechanical behaviour of granite". In: Kolymbas D (ed) Constitutive modelling of granular materials 2000; Springer, Berlin, pp 397–411

[47] Haase R. 1990, Thermodynamics of Irreversible Processes. Dover Publications, New York.

[48] Willam KJ and Warnke EP. Constitutive models for the triaxial behavior of concrete. Proceedings of the International Assoc. for Bridge and Structural Engineering 1975; vol 19, pp. 1- 30.

[49] Ciantia MO, Castellanza R and di Prisco C. Experimental study on the water-induced weakening of calcarenites, Rock Mech. Rock Eng. 2014; DOI 10.1007/s00603-014-0603-z

[50] Korelc J. Multi-language and multi-environment generation of nonlinear finite element codes, Engineering with computers 2002; 18(4): 312-27.

[51] Korelc J. Automation of primal and sensitivity analysis of transient coupled problems, Computational mechanics 2009; 44(5):631-49.

[52] Preisig G, Eberhardt E, Smithyman M, Preh A, Bonzanigo L. Hydromechanical rock mass fatigue in deep-seated landslides accompanying seasonal variations in pore pressures, Rock Mechanics and Rock Engineering 2016; 49(6), 2333-51.

[53] Bonzanigo L, Eberhardt E, Loew S. Long-term investigation of a deep-seated creeping landslide in crystalline rock. Part I. Geological and hydromechanical factors controlling the Campo Vallemaggia landslide. Canadian Geotechnical Journal 2007; 44(10), 1157-80.

[54] Cecinato F, Zervos A, Veveakis E. A thermo-mechanical model for the catastrophic collapse of large landslides. International Journal for Numerical and Analytical Methods in Geomechanics 2011; 35(14), 1507-35.

[55] Muir Wood D. Soil Behaviour and Critical State Soil Mechanics. Cambridge University Press: New York, 1990.

[56] Daouadji A and Hicher PY. An enhanced constitutive model for crushable granular materials, International journal for numerical and analytical methods in geomechanics 2010; 34(6), 555-580.